\RequirePackage{tikz}

\documentclass[pdflatex,sn-mathphys]{sn-jnl}

\jyear{2021}%

\theoremstyle{thmstyleone}%
\theoremstyle{thmstyletwo}%

\theoremstyle{thmstylethree}%
\newtheorem{definition}{Definition}%

\usepackage{amsmath}
\usepackage[caption=false,font=footnotesize]{subfig}
\usepackage{url}
\usepackage{verbatim}
\usepackage{bbm}
\usepackage{color}

\definecolor{lg}{gray}{0.90}
\usepackage{booktabs}
\usepackage{blindtext}
\usepackage{listings}
\usepackage{marvosym}

\graphicspath{{.}}
\usepackage[most]{tcolorbox}
\newcommand*{\ClipSep}{0.2cm}
\usepackage[framemethod=TikZ]{mdframed}

\mdfdefinestyle{myfigbox}{innerlinewidth=0.1pt,roundcorner=6pt,linecolor=black,innerleftmargin=0pt,
innerrightmargin=0pt,innertopmargin=0pt,innerbottommargin=0pt}

\usepackage{xspace}
\usepackage[inline]{enumitem}
\usepackage{amssymb}
\usepackage{float}
\usepackage{mathtools}
\usepackage{pifont}
\usepackage{makecell}
\usepackage{totcount}
\usepackage{tabularx}
\usepackage{xcolor,colortbl}
\usepackage{multirow}
\usepackage{hhline}
\usepackage{threeparttable}
\usepackage{rotating}
\usepackage{longtable}

\newcommand{\out}[1]{}

\newcolumntype{"}{@{\hskip\tabcolsep\vrule width 1pt\hskip\tabcolsep}}

\newcommand{\thickhline}{%
    \noalign {\ifnum 0=`}\fi \hrule height 1pt
    \futurelet \reserved@a \@xhline
}

\begin{document}

\title[Computationally Efficient Labeling of Forum Posts]{Computationally Efficient Labeling of Cancer Related Forum Posts by Non-Clinical Text Information Retrieval}

\author[1,2]{\fnm{Jimmi} \sur{Agerskov}}\email{11450@post.au.dk}
\equalcont{These authors contributed equally to this work.}

\author[2,3]{\fnm{Kristian} \sur{Nielsen}}\email{11633@post.au.dk}
\equalcont{These authors contributed equally to this work.}

\author*[1,2]{\fnm{Christian Marius} \sur{Lillelund \url{https://orcid.org/0000-0002-2520-3774}}}\email{cl@ece.au.dk}

\author[1,2]{\fnm{Christian Fischer} \sur{Pedersen \url{https://orcid.org/0000-0002-1266-5857}}}\email{cfp@ece.au.dk}

\affil[1]{\orgdiv{Department of Electrical and Computer Engineering}, \orgname{Aarhus University}, \orgaddress{\street{Finlandsgade 22}, \postcode{DK-8200}, \city{Aarhus N}, \country{Denmark}}}

\abstract{An abundance of information about cancer exists online, but categorizing and extracting useful information from it is difficult. Almost all research within healthcare data processing is concerned with formal clinical data, but there is valuable information in non-clinical data too. The present study combines methods within distributed computing, text retrieval, clustering, and classification into a coherent and computationally efficient system, that can clarify cancer patient trajectories based on non-clinical and freely available information. We produce a fully-functional prototype that can retrieve, cluster and present information about cancer trajectories from non-clinical forum posts. We evaluate three clustering algorithms (MR-DBSCAN, DBSCAN, and HDBSCAN) and compare them in terms of Adjusted Rand Index and total run time as a function of the number of posts retrieved and the neighborhood radius. Clustering results show that neighborhood radius has the most significant impact on clustering performance. For small values, the data set is split accordingly, but high values produce a large number of possible partitions and searching for the best partition is hereby time-consuming. With a proper estimated radius, MR-DBSCAN can cluster 50000 forum posts in 46.1 seconds, compared to DBSCAN (143.4) and HDBSCAN (282.3). We conduct an interview with the Danish Cancer Society and present our software prototype. The organization sees a potential in software that can democratize online information about cancer and foresee that such systems will be required in the future.}


\maketitle

\section{Introduction}\label{sec:introduction}

The three predominating types of diseases in today's world are cancers, respiratory disorders, and cardiovascular diseases. These diseases entail a predictable trajectory for the patients, caretakers, and relatives, which can be summarized as \cite{jensen_2017_1, murray_2005_1}:  \\

\begin{mdframed}[roundcorner=5pt]
\centerline{ Symptoms ~$\rightarrow$~ Diagnosis ~$\rightarrow$~ Treatment ~$\rightarrow$~ Outcome}
\end{mdframed}
 
Each of the four sequential steps are however complex and encompass a range of concerns, for example life expectancy, patterns of decline, probable interactions with health related services, treatment plans, medical side effects, palliative care, and more. It is difficult, but important to clarify and communicate these complex trajectories to the patient, caretakers, and the relatives. Better informed individuals have better treatment outcomes due to proper disease management and democratized trajectories can lead to better clinical decisions being made, fewer side effects of treatments and fewer re-admissions. Improving the overall care can be done by estimating, clarifying and communicating patient-specific disease trajectories.

Relevant but unused information about diseases and continuity of care is freely available in online communities and forum posts. This information could benefit cancer patients, relatives, or caretakers. Approximately one third of the world's population receive a cancer diagnosis during their lifetime \cite{kraeftens_bekaempelse}. This creates a large field of potential users that would like to learn more about their diagnosis from others. A cancer diagnosis leads to many different reactions, but most tend to seek information online about the trajectory prognosis. A popular trend is to write and communicate on online forums about health issues \cite{Umefjord01042006,medical2016consulations}. On such forums, people write freely on a given topic. On cancer forums, people usually write about their frustrations, experiences, emotions, feelings, and personal preferences regarding any cancer related topic. The established health care systems do not leverage all of this non-clinical information. Some of it may be of clinical relevance, and some only of personal relevance. However, both kinds of information can empower patients, caretakers and relatives by for example strengthening their understanding of a cancer diagnosis, build self-confidence, and establish online or physical communities.

The objective of this present study is to clarify and communicate cancer patient trajectories by information retrieval and subsequent clustering using three common techniques: MR-DBSCAN, DBSCAN, and HDBSCAN. The methods are evaluated on training sets of various sizes (5000 to 25000 posts) in terms of computational efficiency and Rand measure. To our knowledge, this study is the first to perform text retrieval, clustering and classification of cancer patient trajectories in non-clinical texts. The end result is a software prototype that can sift, filter and present cancer information in a visually appealing manner, as we demonstrate with a graphical user-inference. We als oconducted an interview with the Danish Cancer Society \cite{kraeftens_bekaempelse}, who saw a great potential in the presented software prototype and stressed the importance of patient-empowerment. Currently, the digital adoption rate for elderly people in Europe and the US is generally low, but digital maturity is expected to increase in the coming years.

\subsection{Related Work}\label{sec:relatedwork}

Existing research with various objectives, methods and data backgrounds have been addressing the idea of mining data for clarifying and estimating disease trajectories; e.g., natural language processing has a transformative potential within this area (e.g.~\cite{Wang2018,Luque2019,Simpson2012}). In 2005, Murray et
al.~\cite{murray_2005_1} carried out a clinical review that describes
three typical disease trajectories, namely: cancer, organ failure
(heart and lung focus) and frail elderly (dementia focus) and in 2008
Meystre et al.~\cite{Meystre2008} did a review on research within
information extraction from clinical notes in narrative style. Studies
have shown, that even for data normally viewed as highly distinct,
e.g., lab records, a portion of relevant information may only be
available as part of clinical text~\cite{Demner-Fushman2009}. In a
2010 study, Ebadollahi et al.~\cite{ebadollahi_2010_1} predict patient
trajectories from temporal physiological data, and in a 2014 study by
Jensen et al.~\cite{jensen_2014_1} disease observations across a span
of fifteen years from a large patient population were translated into
disease trajectories. In 2016 Ji et al.~\cite{ji_2016_1} developed
prediction models for health condition trajectories and co-morbidity
relationships based on social health records, and in 2017 Jensen et
al.~\cite{jensen_2017_1} conducted a text mining study on electronic
health records in order to automatically identify cancer patient
trajectories. In 2019 Assale et al.~\cite{assale2019} reviewed and
documented the potential of leveraging the unstructured content in
electronic health records. In 2021 Nehme et al.~\cite{Nehme2021} did a
study on natural language processing in the domain of gastroenterology
primarily focusing on structured text within endoscopy, inflammatory
bowel disease, pancreaticobiliary, and liver diseases.

None of the above mentioned studies deal with text
information retrieval, distributed clustering, and classification for
identifying cancer patient trajectories from non-clinical texts,
i.e. online forum posts.

Frunza et al.~\cite{Frunza:2011:MLA} did a related study in 2011; in
their study, they automatically extract sentences from clinical papers
about diseases and/or treatments. Based on the extracted sentences,
semantic relations between diseases and associated treatments are then
identified. Another related study was done by Rosario et
al.~\cite{Rosario:2004:CSR:1218955.1219010} in 2004. The focus of
their work was to recognize text-entities containing information about
diseases and treatments. They use Hidden Markov Models and Maximum
Entropy Models to perform the entity and disease-treatment
relationship recognition. Compared to the Frunza et
al.~\cite{Frunza:2011:MLA} and the Rosario et
al. studies~\cite{Rosario:2004:CSR:1218955.1219010} that focus mostly
on classification, the present study focuses also on text retrieval
and clustering. Further, the present study focuses especially on
cancer trajectories where the other studies have a broader perspective
and aim to cover diseases in general.

In the 2011 study by Yang et al.~\cite{yang2011analyzing},
Density-Based Clustering was used to identify topics within online
forum threads on social media. They also developed a visualization
tool to provide an overview over the identified topics. The purpose of
their tool was to extract topics with sensitive information related to
terrorism or other crime activities; however, it might also be
tailored to extract other topics. Besides using DBSCAN, the study
proposed a related clustering method, namely SDC (Scalable Density
based Clustering). The structure of the Yang et al. study is, to some
extent, similar to the present study; specifically, in the present
study, topics are also extracted from online forum posts, density
based clustering is also used, and result visualization capabilities
are also provided.

\subsection{Reading guide}

Section \ref{sec:system_architecture} presents the overall system architecture. Section \ref{sec:text_retrieval} presents text information retrieval methods. Section \ref{sec:clustering} presents the clustering techniques used in this study. In section \ref{sec:classification}, we detail how further filtering of the clusters are performed as we split them into five categories (Cure, No cure, Disease, Treatment, Side effect and Irrelevant). The results and discussion are presented in sections \ref{sec:results} and \ref{sec:discussion}, respectively.

\section{System Architecture}
\label{sec:system_architecture}

\subsection{Overview}

\begin{figure}[htbp!]
\begin{centering}
  \includegraphics[width=0.54\columnwidth]{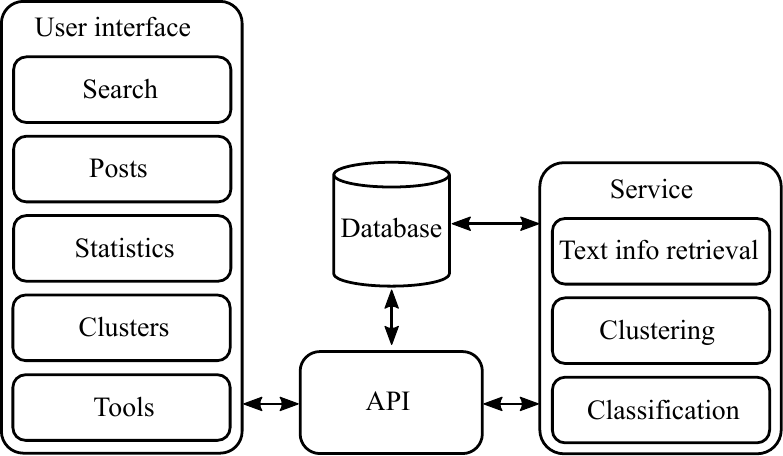}
\caption{Overall component diagram of the implemented software system.}
\label{fig:component_diagram}
\end{centering}
\end{figure}

\begin{figure}[htbp!]
\begin{centering}
  \includegraphics[width=0.78\columnwidth]{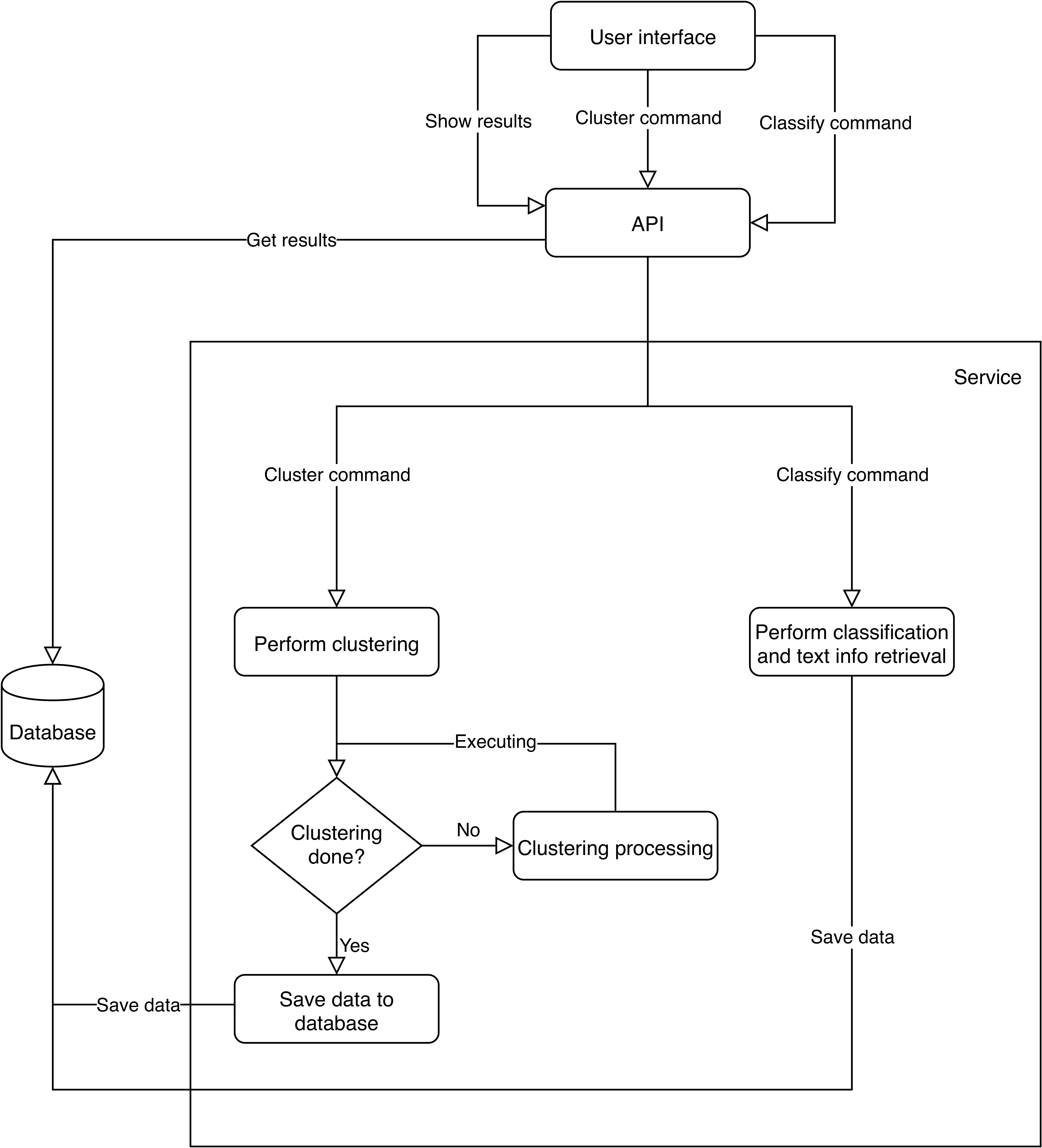}
\caption{Overall activity diagram of the implemented software system.}
\label{fig:activity_diagram}
\end{centering}
\end{figure}

The present study's developed software solution consists of four
components including a database component for storage of clusters. The
solution has been designed in a micro-service architecture with one
process per component. Figure~\ref{fig:component_diagram} provides a
static overview in terms of a component diagram and
figure~\ref{fig:activity_diagram} provides a dynamic overview in terms
of an activity diagram.

The \textit{User interface (UI)} component handles end-user
interaction; this is detailed further in
section~\ref{subsec:user_interface}. The purpose of the \textit{API}
component is primarily to enable easy access for the UI to the
\textit{Database} and the \textit{Service} components. The
\textit{Database} persists all gathered forum posts and the computed
results, e.g. clusters, classes and cancer-trajectories. The
\textit{Service} component is handling the computationally burdensome
data processing; the micro-service architecture enables scaling of
this component only. By implementing the service component as a
scalable unit, it becomes well-suited for the application of a
distributed computing approach. Especially the clustering calculations
are burdensome and needs to be made efficient. Currently, the text
retrieval and classification calculations do not need to be scaled as
they are much faster than the clustering.

\subsection{User Interface}
\label{subsec:user_interface}

Having a tool to visualize data is helpful for effective exploration
of results. The developed user interface is useful for exploring the
collected data set of forum posts and to show information from an area
of interest. For instance, a user is able to select a cluster, i.e. a
cancer-type, of interest, e.g. \textit{breast cancer}, and only
receive posts within that particular cluster. In addition, a user can
also choose a class-label, e.g. \textit{side effect}, and thereby see
all posts from the \textit{breast cancer} cluster that contains
information about \textit{side effects}. Such a tool is both relevant
for scientific use and for cancer patients and caretakers.

\begin{figure}[htbp]
    \centering
    \begin{mdframed}[style=myfigbox,userdefinedwidth=0.75\textwidth,align=center]
    \tcbincludegraphics[blank,arc=\ClipSep]{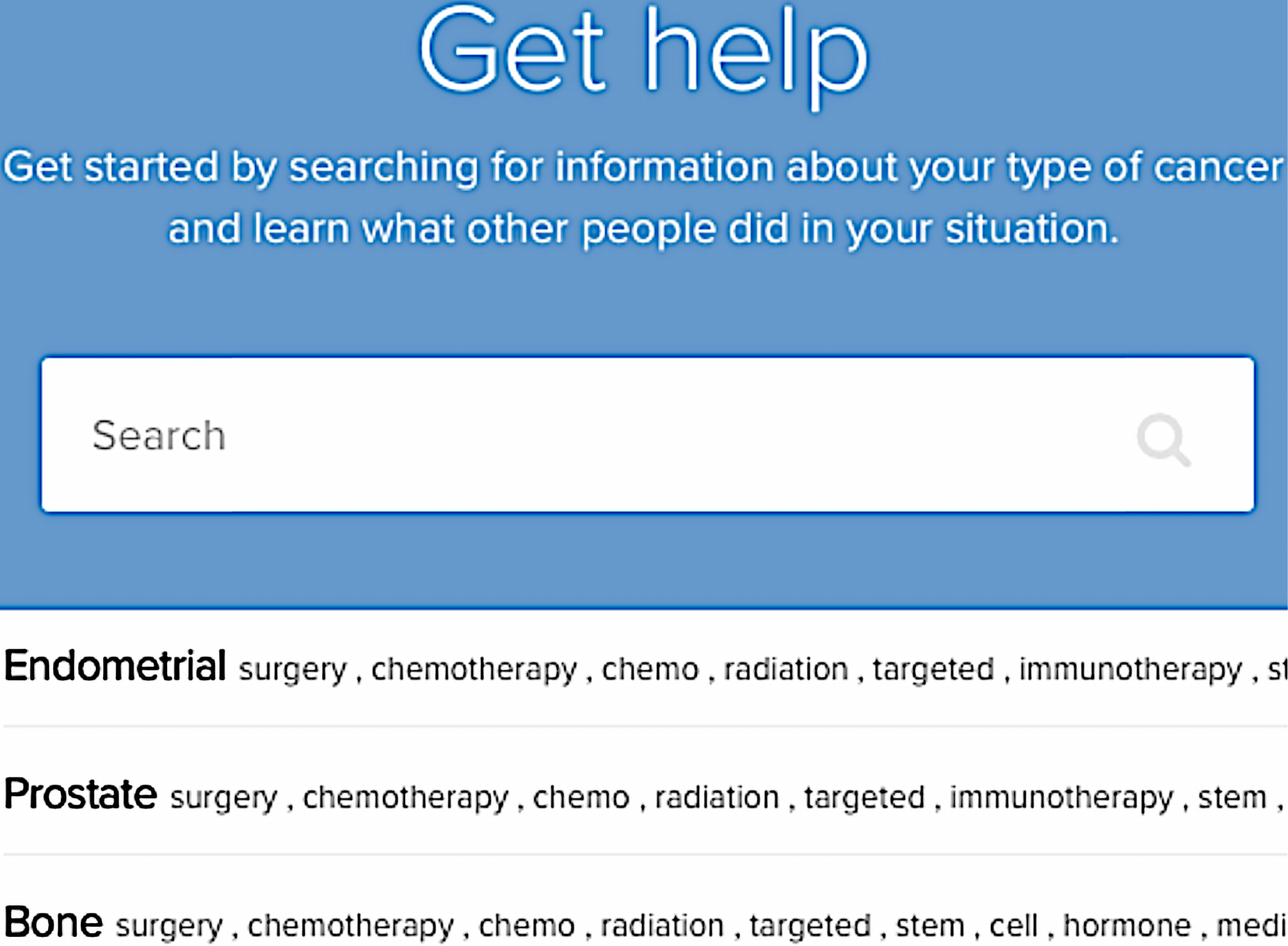}
    \end{mdframed}
    \caption{\textit{Search} view. The cancer-types
    and their associated treatments that have been identified in the
    collection of forum posts are displayed.}
    \label{fig:ui_search}
\end{figure}

\begin{figure}[htbp]
    \centering
    \begin{mdframed}[style=myfigbox,userdefinedwidth=0.75\textwidth,align=center]
        \tcbincludegraphics[blank,arc=\ClipSep]{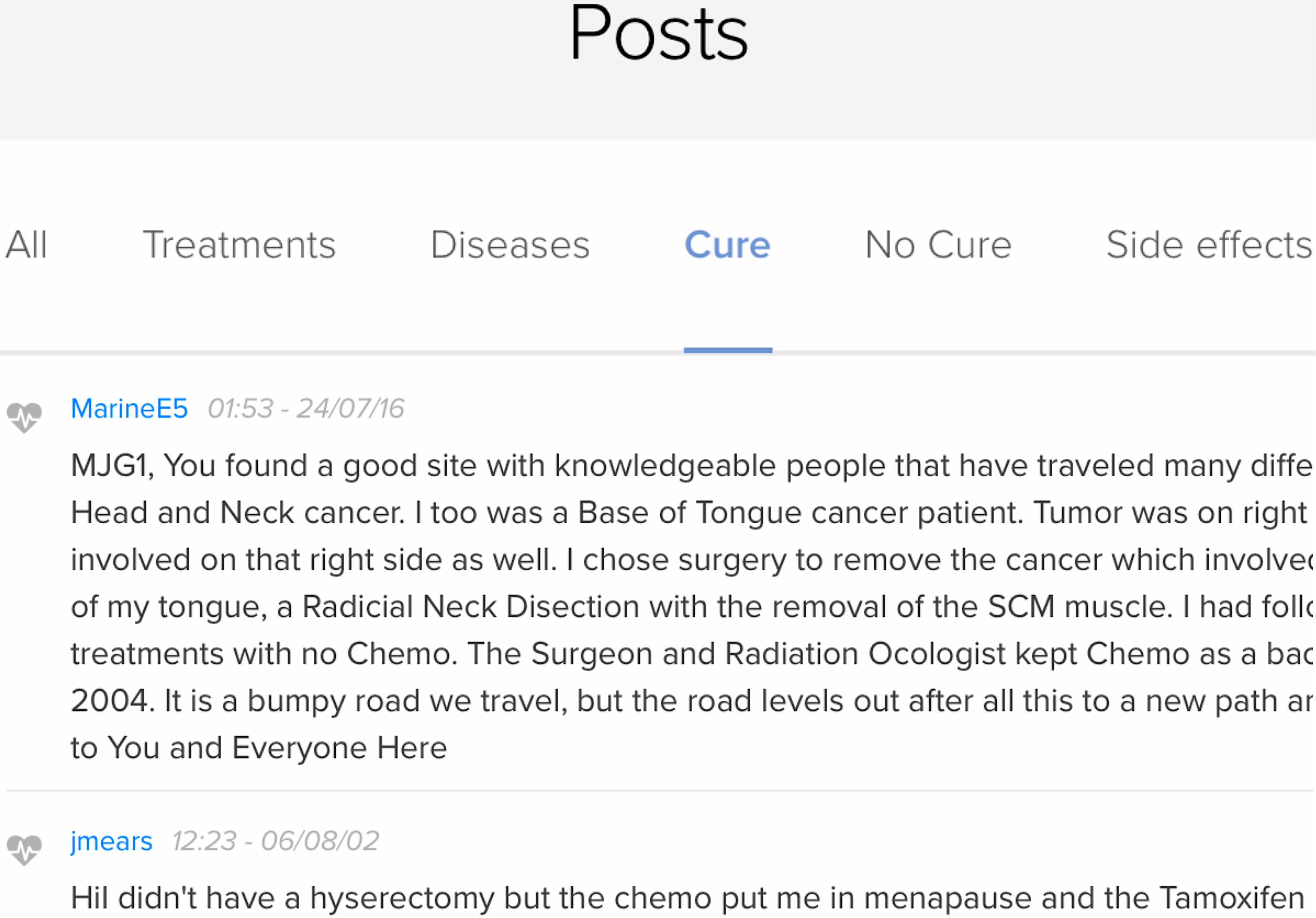}
    \end{mdframed}
  \caption{\textit{Posts} view. All posts
    associated with a selected cancer-type are displayed, and by
    selecting a class-label, i.e. \textit{Disease},
    \textit{Treatment}, \textit{Side effect}, \textit{Cure}, or
    \textit{No cure}, the displayed set of posts are further
    refined.}
    \label{fig:ui_posts}
\end{figure}

\begin{figure}[htbp]
    \centering
  \begin{mdframed}[style=myfigbox,userdefinedwidth=0.75\textwidth,align=center]
    \tcbincludegraphics[blank,arc=\ClipSep]{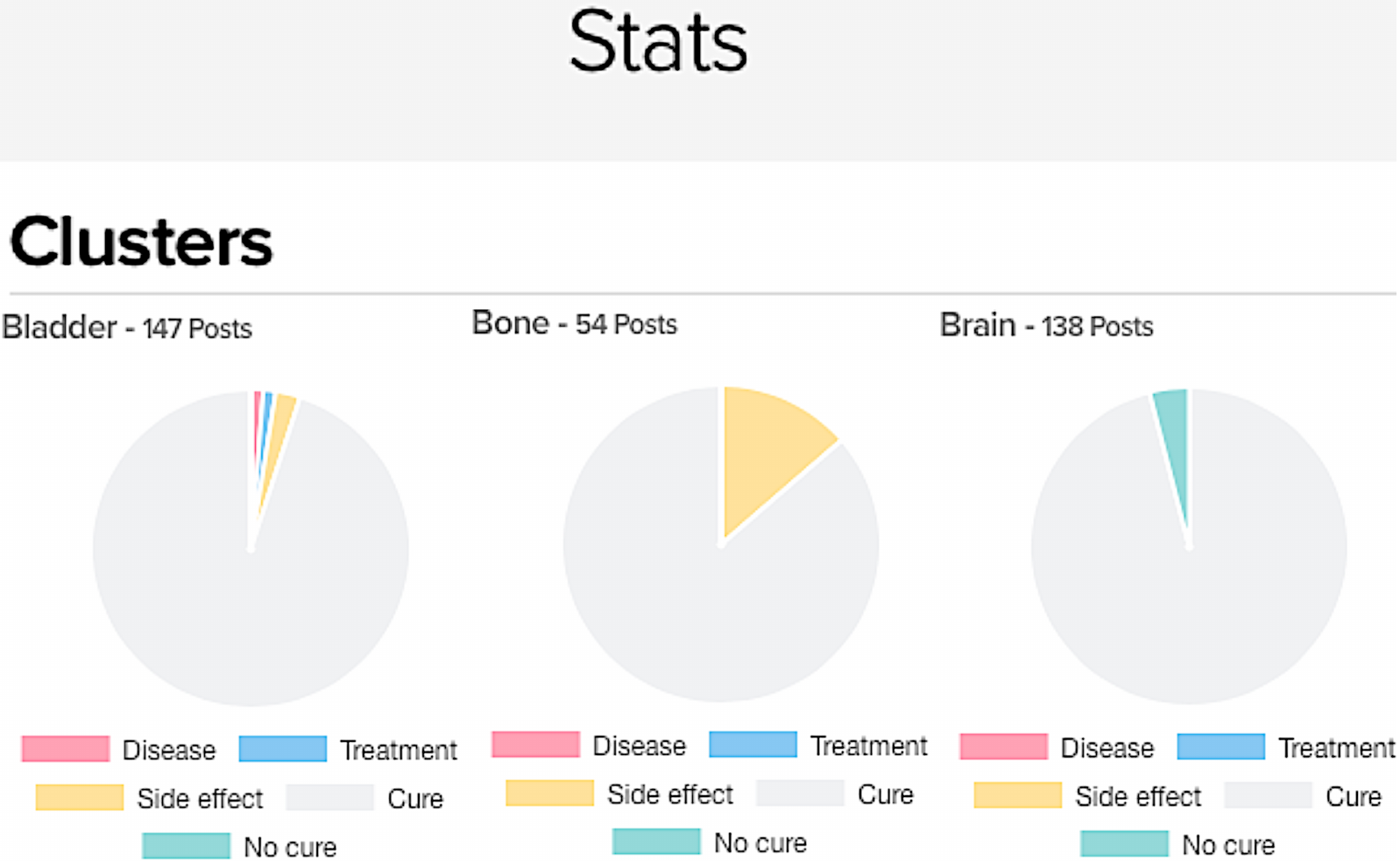}
  \end{mdframed}
  \caption{\textit{Statistics} view. A range of
    relevant descriptive statistics about the posts and their contents
    are displayed in this view. For instance, cancer-type clusters are
    displayed along with their class-label, i.e. \textit{Disease},
    \textit{Treatment}, \textit{Side effect}, \textit{Cure}, or
    \textit{No cure}, distributions.}
    \label{fig:ui_stats}
\end{figure}

The user interface consists of five main views, namely
\textit{Search}, \textit{Posts}, \textit{Statistics},
\textit{Clusters}, and \textit{Tools}
(figure~\ref{fig:component_diagram}). Edited excerpts from the views
\textit{Search}, \textit{Posts} and \textit{Statistics} are seen in
figures~\ref{fig:ui_search},~\ref{fig:ui_posts} and~\ref{fig:ui_stats}
respectively. In the \textit{Search} view, a user can search the
entire collection of forum posts; the identified clusters,
i.e. cancer-types, are displayed along with treatments mentioned in
the posts. By clicking a cancer-type cluster, all posts associated
with that particular cancer-type cluster are displayed in the
\textit{Posts} view. Users can browse through the posts within a
cancer-type cluster, and by selecting a class-label,
i.e. \textit{Disease}, \textit{Treatment}, \textit{Side effect},
\textit{Cure}, or \textit{No cure}, only the posts within the
cancer-type cluster and with the selected label are displayed. In the
\textit{Statistics} view, all cancer-type clusters are displayed along
with their class-label distributions. Also, a histogram showing posts
per cancer-type cluster is displayed along with absolute counts of
posts, clustered posts, clusters, and class-labels. Thereby, the
\textit{Statistics} view provides a useful overview for the end-user;
such an overview is very hard to obtain for any regular end-user
reading through forum posts.

\subsection{Validation}

In this study, all developed software has been evaluated against five out of the eight properties defined in the software product quality model specified in the ISO 25010 standard \cite{iso_25010}. Concretely, the validation has focused on the following properties and sub-properties:

\begin{enumerate}
\item Functional suitability:
  \begin{enumerate*}
  \item Appropriateness,
  \item Correctness
  \end{enumerate*}
\item Performance:
  \begin{enumerate*}
  \item Time behavior,
  \item Resource utilization
  \end{enumerate*}
\item Compatibility:
  \begin{enumerate*}
  \item Interoperability,
  \item Co-existence
  \end{enumerate*}
\item Maintainability:
  \begin{enumerate*}
  \item Modularity,
  \item Modifiability
  \end{enumerate*}
\item Portability: 
  \begin{enumerate*}
  \item Installability
  \end{enumerate*}
\end{enumerate}

\noindent All properties and sub-properties were met by the present study's
software.

\section{Text Information Retrieval}
\label{sec:text_retrieval}

\subsection{Data Collection}

The data consists of automatically collected posts from a set of
publicly available cancer related forums; the posts are written by
medical laypersons. Typically, the posts contain some combination of
diagnoses, symptoms, experiences, questions, side-effects, treatments
and/or treatment outcomes. In this study, each post is saved in the
following self explanatory structure:\newline

\centerline{[ \textit{thread\_id, author, title, date, content} ]}
\vspace{\baselineskip}

The most interesting piece of information in each post is stored in
the \textit{content} attribute. This attribute contains all of a
post's text, and based on this text, information retrieval is done and
relevant features for the cancer-type clustering are extracted. Often,
the post-texts contain rather detailed descriptions of a particular
cancer-type and a received treatment. For example, a user on Cancer
Survivors Network~\cite{csn_cancer} wrote the following about thyroid
cancer:

\begin{quote}\interlinepenalty=10000
  \textit{Hi, everyone I was diagnosed with Papillary Thyroid Caner a little
  more than a year ago. It was right before Christmas last year and I
  went in to get all of my thyroid removed. It was a 5 hour surgery
  and I stayed in the hospital for 3 days because of complications
  with my levels. But, they didn't look at the lymph nodes and the
  last two ultrasounds I have had revealed lymph nodes in my neck and
  one in my throat that have gotten bigger. I need some advice for
  thos who have had their thyroid cancer come back. My treatment of
  thyrogen shots, lab work, and scans start April 1st, 2013. I know
  this cancer is the easiest to treat but I'm wondering what happens
  if it has in fact spread to my lymph nodes and in my body? Does that
  change the prognosis or treatment or staging? Thanks for the help!}
\end{quote}

Although this representative example text is written by a layperson, it
still contains relevant health and cancer related information that
might be useful for other patients or caretakers.

\subsection{Text Retrieval Preprocessing}

In order to perform the actual text information retrieval
successfully, the text needs to be preprocessed. In this study, we
have conducted three preprocessing steps:
\begin{enumerate*}
\item cleansing,
\item stemming, and
\item tokenization.
\end{enumerate*}

In the \textit{cleansing} part, unwanted characters, e.g. HTML tags,
emojis and ASCII-artwork, are removed. This is a non-trivial task when
dealing with forum posts as people express themselves quite
informally.

In the \textit{stemming} part, inflected and derived words are reduced
to their word
stem~\cite{Introduction_to_information_retrieval}. Several different
algorithms for stemming exists, e.g. the \textit{Lovins
  Stemmer}~\cite{lovins1968development}, the \textit{Paice
  Stemmer}~\cite{chris1990another}, and the predominant \textit{Porter
  Stemmer}~\cite{porter1980algorithm}. All of these stemming
algorithms are best suited for English; in the present study, the
Porter Stemmer is used. The Porter Stemming algorithm is based on five
steps, and in each step, a specified set of rules are applied to the
word being processed; for instance, table \ref{porter_step_1a_rule}
shows the processing rules of the first
step~\cite{porter1980algorithm}.

\begin{table}[htbp!]
\setlength\extrarowheight{1pt}
\caption{Exemplification of some of the processing rules in the Porter Stemming algorithm.}
\label{porter_step_1a_rule}
\centering
\begin{tabular}{ |c c c| c |c c c| }
  \cline{1-3} \cline{5-7}
 \multicolumn{3}{|c|}{Rules} &~  & \multicolumn{3}{|c|}{Examples} \\ \cline{1-3} \cline{5-7}
 sses & $\rightarrow$ & ss &~  & care\underline{sses} & $\rightarrow$ & care\underline{ss}\\
 ies    & $\rightarrow$ & i     &~ & difficult\underline{ies} & $\rightarrow$ & difficult\underline{i}\\
 ss     & $\rightarrow$ & ss  && care\underline{ss} &$\rightarrow$ & care\underline{ss}\\
 s       & $\rightarrow$ & \textvisiblespace && treatment\underline{s} & $\rightarrow$ & treatment\textvisiblespace \\\cline{1-3} \cline{5-7}
\end{tabular}
\end{table}

In the \textit{tokenization} part, character and word sequences are
sliced into tokens. Typically, the tokens are words or terms, but in
this study, tokens are only words. After the tokenization, stop words
are removed.

\subsection{Text Retrieval}

For the subsequent clustering of posts into cancer-type clusters to be
accurate, information from all the posts' \textit{content} attribute
needs to retrieved. This is done by using text retrieval together with
a predefined feature vector containing names of a range of cancer
types.

In this study, we use the \textit{Term Weighting} approach. This
approach uses \textit{Term Frequency} and \textit{Inverse Document
  Frequency} to yield \textit{Term Frequency - Inverse Document
  Frequency} which is the final weight of a term.

The purpose of Term Frequency (tf) is to measure how often a term
occurs in a \textit{specific} document, i.e. in this study tf is
simply an unadjusted count of term appearances.

\begin{definition}Term Frequency~\cite{simple-proven-approaches-to-text-retrieval}.\\
  tf$(t,d) \equiv$ occurrences of term $t$ in document $d$. \hfill $\square$
\end{definition}

Clearly, documents vary in length which entails a bias in tf; that is,
a term is likely to appear more often in a long document than in a
short document, given the documents are similar in
content~\cite{On_term_selection_for_query_expansion}. Whenever a term
is frequent in a document it is likely to be important to that
\textit{specific} document.

The purpose of Inverse Document Frequency (idf) is to measure the
weight of a term in a \textit{collection} of documents; a rare term is
often more valuable than a frequent term in a \textit{collection} of
documents~\cite{UnderstandingInverseDocument}.

\begin{definition}Inverse Document Frequency~\cite{simple-proven-approaches-to-text-retrieval}.\\
  idf$(t,D) \equiv  \log{(N)} - \log{(n)}$.\\
  Where $N$ is the number of documents in collection $D$, and $n$ is
  the number of documents in $D$ in which term $t$ appears. \hfill
  $\square$
\end{definition}

Term Frequency - Inverse Document Frequency (tf-idf) is a measure of
how important a word is to a \textit{specific} document in a
\textit{collection} of documents. A large tf-idf weight is obtained
whenever:
\begin{enumerate*}
\item the term frequency is high for the \textit{specific} document, and
\item the document frequency is low for the term across the \textit{collection} of documents.
\end{enumerate*}
The combination of the $tf$ and $idf$ weights tend to filter out
common terms that do not carry much information.

\begin{definition}Term Freq. - Inverse Document Freq.~\cite{A_statistical_interpretation_of_term_specificity}.\\
tf-idf$(t,d,D) \equiv$ tf$(t,d) \cdot$ idf$(t,D)$. \hfill $\square$
\end{definition}

\section{Clustering}
\label{sec:clustering}

\subsection{DBSCAN Clustering}

Clustering is the process of splitting an unlabeled data set into
clusters of observations with similar traits such that intra-cluster
variation is minimized and inter-cluster variation is maximized. A
cluster in this study is a collection of similar posts in terms of
cancer-type.

Density-Based Spatial Clustering of Applications with Noise (DBSCAN)
is a clustering algorithm based on the density of data points (also
known as observations). It creates clusters from regions that have a
sufficiently high density of data points and in doing so, it allows
clusters of any shape even if it contains noise or outliers. This
allows DBSCAN to create non-convex and non-linearly separated
clusters, contrary to many other clustering
algorithms~\cite{Jain2010651, Campello2013}. Also, the algorithm is
able to find clusters of arbitrary
size~\cite{data_mining_concepts,mrdbscan2011} and it does not require
the number of clusters
beforehand~\cite{Ester96adensity_based,data_mining_concepts,mrdbscan2011}. Moreover,
DBSCAN (variants thereof) is horisontally scalable such that efficient
computing can be achieved via a distributed or cluster computing setup
which is attractive, and in some cases even necessary, for large scale
text data processing. Before outlining the DBSCAN algorithm, a number
of associated definitions need to be in place.

\begin{definition}DBSCAN related definitions~\cite{Ester96adensity_based}.\\
\label{def:dbscan}  
  The $\pmb{\varepsilon}$\textbf{-neighborhood} of point $p$
  is defined by the points within a radius $\varepsilon$ of $p$.\\

  If a point $p$'s $\varepsilon$-neighborhood contains at least
  $m_{pts}$ number of points, the point $p$ is called a \textbf{core
    point}.\\

  A point $p$ is \textbf{directly density-reachable} from a point
  $q$ if $p$ is within the $\varepsilon$-neighborhood of $q$ and $q$ is a
  core point.\\

  A point $p$ is \textbf{density-reachable} from a point $q$ with
  regard to $\varepsilon$ and $m_{pts}$ if there is a chain of points,
  $p_1, ..., p_n$, where $p_1 = q$ and $p_n = p$ such that $p_{i+1}$
  is directly density-reachable from $p_i$.\\

  A point $p$ is \textbf{density-connected} to a point $q$ with
  regard to $\varepsilon$ and $m_{pts}$ if there is a point $o$ such that
  both $p$ and $q$ are density-reachable from $o$.\\

  A point $p$ is a \textbf{border point} if $p$'s
  $\varepsilon$-neighborhood contains less than $m_{pts}$ and $p$ is
  directly density-reachable from a core point.\\


  All points not reachable from any other point are outliers called \textbf{noise points}.\\
  
  A \textbf{cluster} $C$ is a non-empty set that satisfies the
  following two conditions for all point pairs $(p, q)$:
  \begin{enumerate*}
  \item if $p$ is in $C$ and $q$ is density-reachable from $p$, then $q$ is also in $C$; and
  \item if $(p,q)$ is in $C$, then $p$ is density-connected to $q$.
  \end{enumerate*}\hfill $\square$
\end{definition}

To establish a cluster, DBSCAN starts with an arbitrary point $p$ and
finds all density-reachable points from $p$ with respect to
$\varepsilon$ and $m_{pts}$. If $p$ is a core point a new cluster with
$p$ as a core point is created. If $p$ is a border point DBSCAN visits
the next point in the data collection. DBSCAN may merge two clusters
into one, if the clusters are
density-reachable~\cite{Ester96adensity_based}. The algorithm
terminates when no new points can be added to any
cluster~\cite{data_mining_concepts}.

\subsection{MR-DBSCAN Clustering}
\label{ch_theory:clustering_mrdbscan}

\begin{figure}[htbp!]
    \centering
    \includegraphics{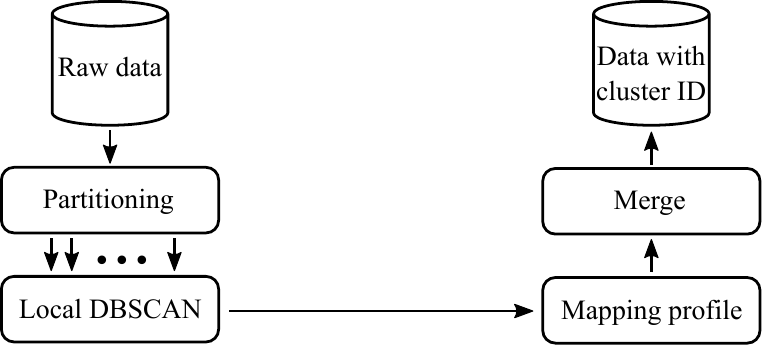}
    \caption{%
      An overview of the steps in MR-DBSCAN. %
      In the \textit{Partitioning} step, the available raw data points
      are split into partitions based on a set of criteria. %
      The \textit{Local DBSCAN} step performs parallel DBSCANs on the
      received partitions and outputs a set of merge candidates. %
      The \textit{Mapping profile} step deals with issues with cross
      border constraints when merging partitions, e.g. a data point in
      one partition might be connected to a cluster in another
      partition. The output is a list of cluster pairs that should be
      merged although they cross bordering partitions. %
      The \textit{Merge} step builds a cluster profile map for mapping
      local cluster IDs into global cluster IDs; hereafter, the
      mapping is done for all data points.%
    }
    \label{fig:mrdbscan_overview}
\end{figure}

Clustering with DBSCAN is computationally burdensome both with regard
to run-time and memory consumption~\cite{xu1999fast}. To achieve
run-time efficiency, MR-DBSCAN (MapReduce-DBSCAN)~\cite{mrdbscan2011,
  mrdbscan:He2014} is used rather than DBSCAN. Besides distributing
computations via MapReduce, the two clustering algorithms are
equivalent. Figure~\ref{fig:mrdbscan_overview} outlines the steps in
MR-DBSCAN.

\subsubsection{Partitioning}

To maximize parallelism and thus the run-time efficiency gain, data
must be well balanced such that data, and thus the computational
work-load, can be evenly distributed on the compute-nodes. However,
data in real life applications are often unbalanced and this needs to
be addressed with a suitable data partitioning strategy; such a
strategy is part of MR-DBSCAN.

One possible partitioning strategy is to recursively split the entire
data set into smaller sets, i.e. partitions, until a stop criterion is
met, e.g. all partitions contain less than a given number of points or
a given number of partitions have been made. According to
definition~\ref{def:dbscan}, the geometry of a cluster, and therefore
sensibly also a partition, cannot be smaller than $2 \varepsilon$, so
when splitting a partition, the geometry must remain extended beyond
$2 \cdot 2 \varepsilon$.  When splitting a partition into two in
MR-DBSCAN~\cite{mrdbscan2011}, all possible splits are considered. The
split that minimizes the cost in one of the sub-partitions are
chosen. Here, \textit{cost} is the difference between the number of points in
sub-partition-1 and half of the number of points in
sub-partition-2. Each partition is given a key and associated with a
reducer.

\subsubsection{Local DBSCAN}
\label{subsec:localdbscan}
\begin{figure}[htbp!]
    \centering
    \includegraphics{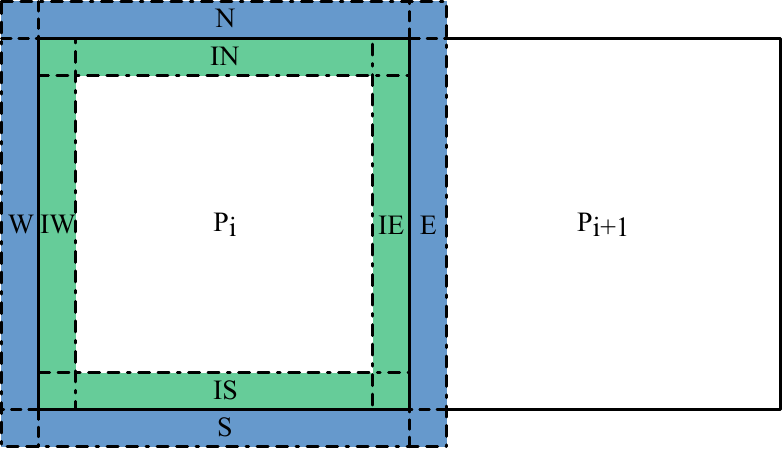}
    \caption{Two MR-DBSCAN bordering partitions $P_i$ and $P_{i+1}$
      along with $P_i$'s extended partition with a blue outer margin
      and a green inner margin (inspired by~\cite{mrdbscan2011}).}
    \label{fig:mrdbscan_margins}
\end{figure}

A reducer is given a partition and all its associated data
points. Therefore, the mapper must prepare all data related to a
partition for every single partition. That is, for instance, a
partition $P_i$, the related data $C_i$ within $P_i$, but also the
data within $P_i$'s $\varepsilon$-width extended partition $R_i$ that
overlap the bordering partitions. In case of a 2D-grid partitioning
those bordering sets are: \textit{North (N, IN)}, \textit{South (S,
  IS)}, \textit{East (E, IE)}, and \textit{West (W, IW)}
(figure~\ref{fig:mrdbscan_margins}).

Local DBSCAN uses the same principles as DBSCAN to perform its
clustering. It starts with an arbitrary data point $p \in C_i$ and
finds all density-reachable points from $p$ with respect to
$\varepsilon$ and $m_{pts}$. If $p$ is a core point, the
$\varepsilon$-neighborhood will be explored. If Local DBSCAN finds a
point in the outer margin that is directly-density-reachable from a
point in the inner margin, it is added to the merge-candidate set. If
a core point is located in the inner margin, it is also added to the
merge-candidate set. Each clustered point is given a local cluster ID,
which is generated from the partition ID and the label ID from the
local clustering: \textit{(partitionID, localclusterID)}. The output
of a reducer is the clustered data points and the merge candidate set.

\subsubsection{Mapping Profile}

After each partition has undergone clustering and merge candidate
lists have been generated, the merge candidate lists are collected to a
single merge candidate list. The basics of merging the clusters from
the different partitions are:
\begin{enumerate*}
\item Execute a nested loop on all points in the collected merge
  candidate lists to see if the same data points exists with different
  local cluster IDs;
\item If found, then merge the clusters.
\end{enumerate*}

\begin{figure}[htbp!]
    \centering
    \includegraphics{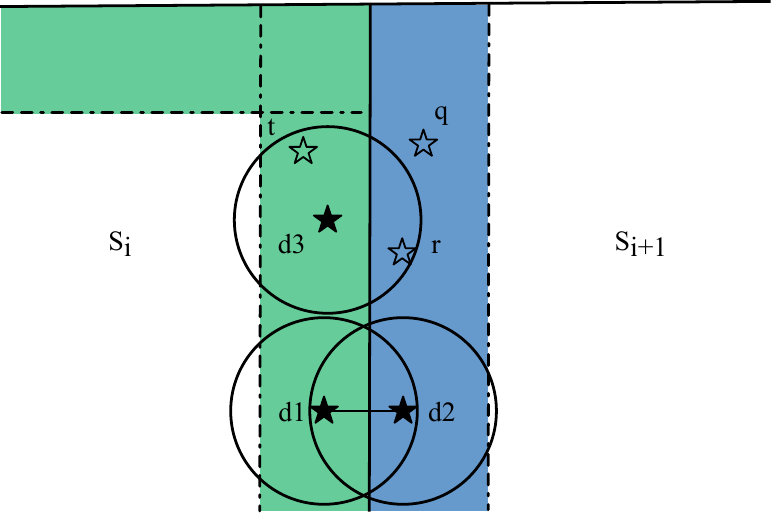}
    \caption{Two MR-DBSCAN bordering partitions $S_i$ and $S_{i+1}$
      along with $S_i$'s extended partition with a blue outer margin
      and a green inner margin. The points $d_1, d_3 \in C_1$ and
      $d_2 \in C_2$ are core points, and $t \in C_1$ and
      $r, q \in C_2$ are border points (inspired
      by~\cite{mrdbscan2011}).}
    \label{fig:mrdbscan_corner}
\end{figure}

Figure~\ref{fig:mrdbscan_corner} illustrates two examples of
cluster-merge propositions. Example 1: the points $d_1 \in C_1$ and
$d_2 \in C_2$ are core points and $d_2$ is directly density-reachable
from $d_1$; thus, $C_1$ should merge with $C_2$. Example 2: The point
$d_3 \in C_1$ is a core point and $r \in C_2$ is a border point; thus,
$C_1$ should not merge with $C_2$.

As it was seen in the Local DBSCAN step
(subsection~\ref{subsec:localdbscan}), the output of each Local DBSCAN
is a merge-candidate list consisting of two types of points, namely:
\begin{enumerate*}
\item the core points in the inner margin, and
\item directly-density-reachable points in the outer margin.
\end{enumerate*}
Clearly, this is suitable for the present \textit{Mapping Profile}
step where the purpose is to create a profile that maps clusters that
should be merged. The algorithm for generating the mapping profile is
shown in the algorithm in figure~\ref{mrdbscan:find_merge_mapper}.

\begin{figure}[htbp!]
\hrulefill
    \begin{algorithmic}[1]
		\For{each $cp \in$ CP} \Comment{Core Pts}
			\For{each $bp \in$ BP} \Comment{Border Pts}
        		\If{$cp.id == bp.id$}
    	        	\State MP.add(($cp.local\_cluster\_id$), \par \hskip20.6mm ($bp.local\_cluster\_id$)) \Comment{Merge Pts}
                    \State BP.delete($bp$)
              	\EndIf
        	\EndFor
		\EndFor
    \State \Return MP, BP
	\end{algorithmic}
\vspace{-2mm}
\hrulefill
\caption{Generate merge mapping profile~\cite{mrdbscan2011}.}
\label{mrdbscan:find_merge_mapper}
\end{figure}

The output of the algorithm is a list of pairs of local clusters to be
merged (denoted MP) and a list of border points (denoted BP); a point
$p$ is at least a border point in a merged cluster (this is taken care
of in the next step).

\subsubsection{Merge}

The previous step resulted in a list of pairs of clusters to be
merged. The IDs of the local clusters should be changed into a unique
global ID after merging. Thus, a global perspective of all local
clusters is build (algorithm in figure~\ref{mrdbscan:generate_global_id}). The
algorithm generates the map:\\
\centerline{ \textit{(partitionID, localclusterID)} $\rightarrow$ \textit{globalclusterID}.}

\noindent Lastly, as mentioned in the previous step, noise points are
set to border points.

\begin{figure}[htbp!] 
\hrulefill
    \begin{algorithmic}[1]
		\For{each element pair $e_i, \!e_j \!\in \!\text{MP}, i \!\neq\! j,$} \Comment{Merge Pts}
        	\If{$e_i, e_j \notin$ L} \Comment{Mapper L}
                \State put $e_i$ and $e_j$ into the same Map Slot in L
                \EndIf
                \If{$e_i \in \text{L} \land e_j \notin \text{L}$}
                \State put $e_j$ into $e_i$'s Map Slot in L
                \EndIf
                \If{$e_i, e_j \in \text{L}$}
                \State if $e_i$ and $e_j$ are not in the same Map Slot
                in L, \par\hskip3.5mm then move the Map Slot 
                with the highest index to \par\hskip3.5mm the Map Slot with the lowest index
                \EndIf
		\EndFor
    \State \Return L
	\end{algorithmic}
\vspace{-2mm}
\hrulefill
\caption{Generate global ID map~\cite{mrdbscan2011}.}
\label{mrdbscan:generate_global_id}
\end{figure}

\section{Classification}
\label{sec:classification}

The result of the clustering is a set of cancer-type clusters. To
enable further filtering possibilities for the end-user, a
within-cluster classification is conducted such that each post within
a cancer-type cluster is labeled with one of the six labels
illustrated in table~\ref{ch_theory:classification:table_labels}. This
allows an end-user to filter the forum posts such that, for instance,
only posts with breast cancer (cluster) treatments (class) are shown.

\begin{table}[htbp!]
\setlength\extrarowheight{1pt}
\caption{Class labels for conducting within-cluster classification of forum posts.}
\label{ch_theory:classification:table_labels}
\centering
\begin{tabular}{|l|l|}
\hline
\textbf{Class label}       & \textbf{Class description with example post in italics} \\ \hline
Cure                     & \makecell[l]{About cancer curing treatments.\\
                                 \textit{After 16 chemo sessions my cancer was gone.}}\\ \hline
No cure               & \makecell[l]{About cancer non-curing treatments.\\
                                 \textit{My husband went through chemo since he had bladder}\\
                                 \textit{cancer. Sadly he passed.}}\\ \hline
Disease               & \makecell[l]{About cancer without mentioning treatments.\\
                                 \textit{I was diagnosed with breast cancer a few weeks ago.}}\\ \hline
Treatment            & \makecell[l]{About treatments without mentioning cancer.\\
                                 \textit{Has anyone tried being treated with Stem Cells? }} \\\hline
Side effect            & \makecell[l]{Side effects of disease or treatment.\\
                                 \textit{The chemo makes it really hard for me to swallow}\\
                                 \textit{and a hard time eating.}}\\ \hline
Irrelevant              & \makecell[l]{About none of the above.\\
                                 \textit{I am so sorry to hear that. Love Lea.}}\\ \hline
\end{tabular}
\end{table}

We have chosen to classify with a Naive Bayes classifier trained with
a manually created training set augmented with the freely available
set from The BioText Project, UC,
Berkeley~\cite{training_set_bio}. The Frunza et al. study also uses a
Naive Bayes classifier with promising
results~\cite{Frunza:2011:MLA}. However, they classified abstracts
from scientific articles which is a somewhat different data-domain
than the present study's non-clinical texts.

The time complexity for training a Naive Bayes classifier is
$\mathcal{O}(np)$, where $n$ is the number of training observations
and $p$ is the number of features; thus, disregarding the constant,
the complexity is in terms of observations $\mathcal{O}(n)$. When
testing, Naive Bayes is also linear which is optimal for a classifier.

\section{Results}
\label{sec:results}

\subsection{Clustering: DBSCAN and MR-DBSCAN Verification}

MR-DBSCAN is a distributed extension of DBSCAN and they use the same
principles for clustering. Thus, given the same input, the two
clustering methods should yield exactly the same output. The results
in this section show that this is indeed the case and we thereby
consider the implementations of MR-DBSCAN and DBSCAN to be verified in
terms of correctness of logical output. The actual implementations do
not share code so it seems fair to disregard the odd risk of having
both implementations wrong in a manner that lead to the same output.

For comparing the clusterings of DBSCAN and MR-DBSCAN, the Adjusted
Rand Index (ARI)~\cite{Hubert1985} is used. The index is a similarity
measure between two clusterings and it is obtained by counting the
number of identical labels assigned to the same clusters vs. the
number of identical labels assigned to different clusters. If the
label assignments coincide fully, the index is 1, and if they do not
coincide at all, the index is 0. If DBSCAN and MR-DBSCAN are
implemented correctly, the ARI must be 1 regardless of:
\begin{enumerate*}
\item the number of points in the data set,
\item the number of partitions in MR-DBSCAN, and 
\item the parameter settings for $\varepsilon$ and $m_{pts}$. 
\end{enumerate*}
In addition, the number of partitions (\#P) in MR-DBSCAN, the coverage
percentage (\%C), and the number of labels (\#L) in DBSCAN and
MR-DBSCAN have been recorded. The results show
(table~\ref{tab:results_exp_1}) that the ARI is 1 in all 18 test
cases; a necessary condition for this to happen, is that both
MR-DSBCAN and DBSCAN yield the same number of labels in all the tests
which is also the case (table~\ref{tab:results_exp_1}).

Also, MR-DBSCAN has been partitioning its data into 3-8 partitions
(table~\ref{tab:results_exp_1}), which means that even though the data
has been split and clustered individually per partition, the merging
works as intended and yields the same clustering as DBSCAN. The
coverage percentage value is also identical for the two clusterings in
all test cases.

\begin{table}[htbp!]
\setlength\extrarowheight{1pt}
\caption{Adjusted Rand Index (ARI) of DBSCAN and MR-DBSCAN in 18 test cases. The columns left to right detail: 1) the number of test case posts, 2-3) the parameter settings for $\varepsilon$ and $m_{pts}$, 4-7) the number of labels (\#L) and the coverage percentage (\%C), 8) the number of partitions (\#P), and 9) the ARI.}
\label{tab:results_exp_1}
\centering
\begin{tabular}{|c|c|c|c|c|c|c|c|c|c|}
\hline
\multirow{2}{*}{\textbf{Posts}}&
\multirow{2}{*}{\boldmath$\varepsilon$}&
\multirow{2}{*}{$\mathbf{m_{pts}}$}&
\multicolumn{2}{c|}{\textbf{DBSCAN}}&
\multicolumn{3}{c|}{\textbf{MR-DBSCAN}}&
\multirow{2}{*}{\textbf{ARI}} \\ \cline{4-8}
            &                 &       & \textbf{\#L} & \textbf{\%C} & \textbf{\#L} & \textbf{\%C} & \textbf{\#P} &  \\ \hline
 15000 & $10^{-3}$  & 5   & 10 & 3.34 & 10 & 3.34 & 8 & 1 \\ \hline
 15000 & $10^{-3}$  & 50  & 2  & 2.99 & 2  & 2.99 & 8 & 1 \\ \hline
 15000 & $10^{-3}$  & 100 & 1  & 2.66 & 1  & 2.66 & 8 & 1 \\ \hline
 15000 & $10^{-2}$  & 5   & 10 & 3.34 & 10 & 3.34 & 7 & 1 \\ \hline
 15000 & $10^{-2}$  & 50  & 2  & 2.99 & 2  & 2.99 & 7 & 1 \\ \hline
 15000 & $10^{-2}$  & 100 & 1  & 2.66 & 1  & 2.66 & 7 & 1 \\ \hline
 15000 & $10^{-1}$  & 5   & 11 & 3.37 & 11 & 3.37 & 3 & 1 \\ \hline
 15000 & $10^{-1}$  & 50  & 2  & 2.99 & 2  & 2.99 & 3 & 1 \\ \hline
 15000 & $10^{-1}$  & 100 & 1  & 2.66 & 1  & 2.66 & 3 & 1 \\ \hline
 25000 & $10^{-3}$ & 5   & 23 & 2.92 & 23 & 2.92 & 7 & 1 \\ \hline
 25000 & $10^{-3}$ & 50  & 2  & 2.37 & 2  & 2.37 & 7 & 1 \\ \hline
 25000 & $10^{-3}$ & 100 & 1  & 2.02 & 1  & 2.02 & 7 & 1 \\ \hline
 25000 & $10^{-2}$ & 5   & 23 & 2.92 & 23 & 2.92 & 6 & 1 \\ \hline
 25000 & $10^{-2}$ & 50  & 2  & 2.37 & 2  & 2.37 & 6 & 1 \\ \hline
 25000 & $10^{-2}$ & 100 & 1  & 2.02 & 1  & 2.02 & 6 & 1 \\ \hline
 25000 & $10^{-1}$ & 5   & 24 & 2.94 & 24 & 2.94 & 3 & 1 \\ \hline
 25000 & $10^{-1}$ & 50  & 2  & 2.37 & 2  & 2.37 & 3 & 1 \\ \hline
 25000 & $10^{-1}$ & 100 & 1  & 2.02 & 1  & 2.02 & 3 & 1 \\ \hline
\end{tabular}
\end{table}

\subsection{Run-Time Analysis: MR-DBSCAN}
\label{subsection:exp_runtimeanalysis}

The purpose of this experiment is to demonstrate the run-time of each
of the MR-DBSCAN steps under variations in:
\begin{enumerate*}
\item the number of forum posts, and
\item the neighborhood radius $\varepsilon$.
\end{enumerate*}
Clearly, these two parameters have the largest influence on the
MR-DBSCAN's run-time. The $\varepsilon$ parameter is used when
partitioning the data set and therefore it has a direct influence on
the beneficial effects of MapReduce.

In all tests, the lower point-count threshold for establishing a core
point, $m_{pts}$, is fixed to 5 points. This is done as the parameter
only has very little run-time influence and this influence is isolated
to the DBSCAN step, i.e. it does not highlight run-time differences
between DBSCAN and MR-DBSCAN.

%
\begin{table}[htbp!]
\setlength\extrarowheight{1pt}
\caption{Run-time measurements of MR-DBSCAN. Columns left to right detail:
  1) the number of posts in the experiment,
  2) the neighborhood radius $\varepsilon$,
  3-6) run-time of each individual step in MR-DBSCAN,
  7) the total run-time, and
  8) the number of data partitions.
  The lower point-count threshold for establishing a core point, $m_{pts}$, is fixed to 5.
  Gray rows indicate discontinued tests as they exceeded our chosen run-time maximum threshold of 16 minutes.}
\label{tbl:experiment4_test_setup}
\centering
\begin{tabular}{|c|c|c|c|c|c|c|c|}
\hline
   \begin{sideways}  \textbf{Posts [}$\mathbf{10^3}$\textbf{] } \end{sideways}%
& \begin{sideways} \boldmath{$\varepsilon$} \textbf{[1]} \end{sideways}%
& \begin{sideways} \textbf{Partition [s]} \end{sideways}%
& \begin{sideways} \textbf{DBSCAN [s] } \end{sideways}%
& \begin{sideways} \textbf{Map [s]} \end{sideways}%
& \begin{sideways} \textbf{Merge [s]} \end{sideways}%
& \begin{sideways} \textbf{Total [s]} \end{sideways}%
& \begin{sideways} \textbf{\#P} \end{sideways}\\ \hline
5 & $1 \! \cdot \! 10^{-0}$     & 0.41  & 3.82   & 0.00 & 0.02 & 11.16  & 1   \\ \hline
5 & $1 \! \cdot \! 10^{-1}$     & 0.62  & 3.79   & 0.00 & 0.02 & 11.35  & 3   \\ \hline
5 & $1 \! \cdot \! 10^{-2}$     & 1.55  & 1.94   & 0.00 & 0.02 & 12.24  & 7   \\ \hline
\rowcolor{lg} 
5 & $1 \! \cdot \! 10^{-3}$     &  -      &  -        & -                   & -       &  -       & -  \\ \hline
\rowcolor{lg} 
5   & $5 \! \cdot \! 10^{-4}$   & -       & -         & -                   & -       & -        & - \\ \hline
\rowcolor{lg} 
5   &  $1 \! \cdot \! 10^{-4}$  & -       & -         & -                   & -       & -        & -  \\ \hline
10 & $1 \! \cdot \! 10^{-0}$   & 0.86   & 10.44 & 0.00 & 0.04 & 18.15  & 1   \\ \hline
10 & $1 \! \cdot \! 10^{-1}$   & 1.31  & 10.35 & 0.00 & 0.05 & 18.78  & 3   \\ \hline
10 & $1 \! \cdot \! 10^{-2}$   & 2.48  & 3.12  & 0.00 & 0.04 & 13.84  & 7   \\ \hline
10 & $1 \! \cdot \! 10^{-3}$   & 50.64  & 2.73 & 0.00 & 0.03 & 61.85  & 8   \\ \hline
10 & $5 \! \cdot \! 10^{-4}$  & 130.68 & 2.75  & 0.00 & 0.03 & 142.29 & 8  \\ \hline
\rowcolor{lg} 
10 & $1 \! \cdot \! 10^{-4}$   & -        & -        & -                     & -      & -        & -  \\ \hline
15 & $1 \! \cdot \! 10^{-0}$   & 1.17   & 21.49  & 0.00 & 0.06 & 31.96  & 1   \\ \hline
15 & $1 \! \cdot \! 10^{-1}$   & 1.86   & 17.86  & 0.00 & 0.07 & 25.74  & 3   \\ \hline
15 & $1 \! \cdot \! 10^{-2}$   & 3.26   & 3.68   & 0.00 & 0.06 & 14.75  & 7   \\ \hline
15 & $1 \! \cdot \! 10^{-3}$  & 66.89  & 3.38   & 0.00 & 0.05 & 78.34  & 8   \\ \hline
15 & $5 \! \cdot \! 10^{-4}$  & 187.76 & 3.39   & 0.00 & 0.05 & 199.11 & 8  \\ \hline
\rowcolor{lg} 
15 & $1 \! \cdot \! 10^{-4}$  & -         & -         & -                   &  -       &  -        & - \\ \hline
20 & $1 \! \cdot \! 10^{-0}$  & 1.65   & 40.17  & 0.00 & 0.09 & 51.23  & 1   \\ \hline
20 & $1 \! \cdot \! 10^{-1}$  & 2.29   & 30.58  & 0.00 & 0.09 & 38.98  & 3   \\ \hline
20 & $1 \! \cdot \! 10^{-2}$  & 3.86   & 4.18   & 0.00 & 0.08 & 15.544  & 6   \\ \hline
20 & $1 \! \cdot \! 10^{-3}$  & 66.51  & 4.63   & 0.00 & 0.07 & 78.39  & 7   \\ \hline
20 & $5 \! \cdot \! 10^{-4}$  & 182.58 & 4.20   & 0.00 & 0.07 & 194.22 & 7   \\ \hline
\rowcolor{lg} 
20 & $1 \! \cdot \! 10^{-4}$  & -        & -           & -                    & -      & -          & - \\ \hline
25 & $1 \! \cdot \! 10^{-0}$  & 2.43   & 105.83 & 0.00 & 0.10 & 127.76 & 1  \\ \hline
25 & $1 \! \cdot \! 10^{-1}$  & 2.89   & 98.96  & 0.00 & 0.09 & 110.63 & 3  \\ \hline
25 & $1 \! \cdot \! 10^{-2}$  & 4.71   & 5.18   & 0.00 & 0.10 & 16.83  & 6   \\ \hline
25 & $1 \! \cdot \! 10^{-3}$  & 78.75  & 5.93   & 0.00 & 0.08 & 90.76  & 7   \\ \hline
25 & $5 \! \cdot \! 10^{-4}$  & 202.71 & 5.85   & 0.00 & 0.09 & 214.82 & 7  \\ \hline
\rowcolor{lg} 
25 & $1 \! \cdot \! 10^{-4}$  & -        &  -       &  -      & -      &  -       & -  \\ \hline
\end{tabular}
\end{table}

\begin{figure}[htbp!]
    \centering
    \includegraphics[width=0.75\columnwidth]{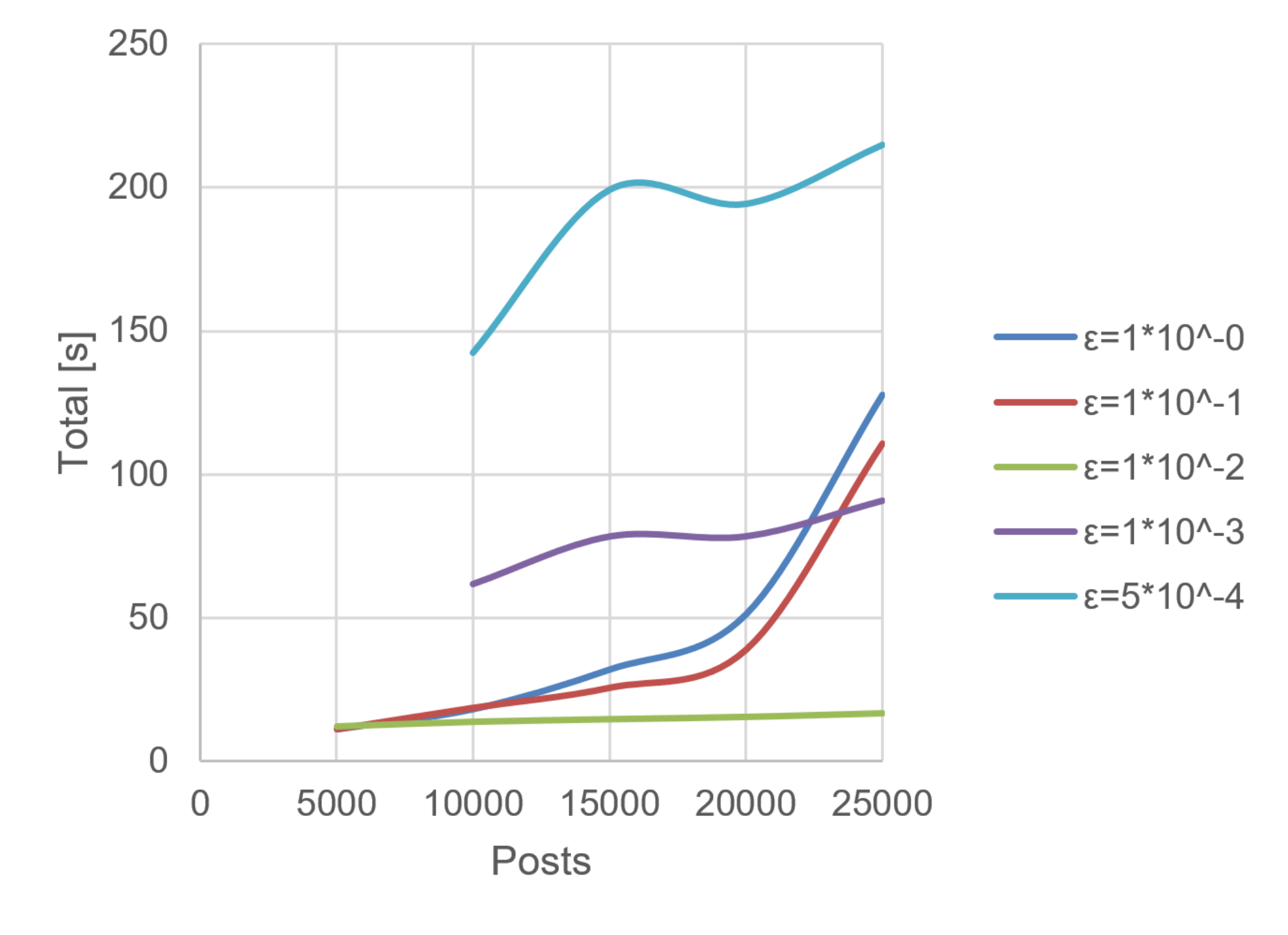}
    \caption{The total run-time is minimized at $\varepsilon = 0.01$
      where the individual run-times for both partitioning and DBSCAN
      is relatively low. At $\varepsilon = 1, 0.1$ and $0.01$
      (relatively large values), the total run-times are dominated by
      DBSCAN. At $\varepsilon = 0.001$ and $0.0005$ (relatively small
      values), the total run-times are dominated by the partitioning.}
    \label{fig:runtimeanalysis_mrdbscan}
  \end{figure}

For all 30 test cases (table~\ref{tbl:experiment4_test_setup}),
mapping takes almost no time at all; merging has also only little
effect on run-time. For relatively large values of $\varepsilon$,
i.e. 1 and 0.1, compared to the data span, MR-DBSCAN is not able to
partition the data set well. Clearly, this affects the run-time as the
clustering then is performed on a single partition (or very few) and
no MapReduce improvements are achieved. For relatively small values of
$\varepsilon$, i.e. 0.001 and 0.0005, the data set is split well into
partitions, but due to the low value of $\varepsilon$ there is a large
number of possible partitions, and a lot of time is spent in search of
the the best partitioning. Thus, as the results show, the partitioning
becomes slower when $\varepsilon$ decreases, but the local DBSCAN
becomes faster. Hence, $\varepsilon$ needs to be set with care to
strike a balance and minimize the total run-time of MR-DBSCAN. In our
experiments, the balance is $\varepsilon = 0.01$
(table~\ref{tbl:experiment4_test_setup} and
figure~\ref{fig:runtimeanalysis_mrdbscan}). At this point, the
partitioning run-time is relatively low and likewise for the local
DBSCAN; this results in a relatively low total run-time.

Decreasing $\varepsilon$ even further to $0.0001$ made the
partitioning exceed the set max limit of total run-time of 16 minutes
(see the grayed-out rows in
table~\ref{tbl:experiment4_test_setup}). Entries are also missing
(table~\ref{tbl:experiment4_test_setup} and
figure~\ref{fig:runtimeanalysis_mrdbscan}) at $5000$ posts and
$\varepsilon = 0.0005$ and $0.001$ as proper equidistributed
partitioning cannot be done when both the number of posts and
$\varepsilon$ are relatively low.

  \subsection{Run-Time Contrast: DBSCAN, MR-DBSCAN, HDBSCAN}

\begin{figure}[htbp!]
    \centering
    \includegraphics[width=0.75\columnwidth]{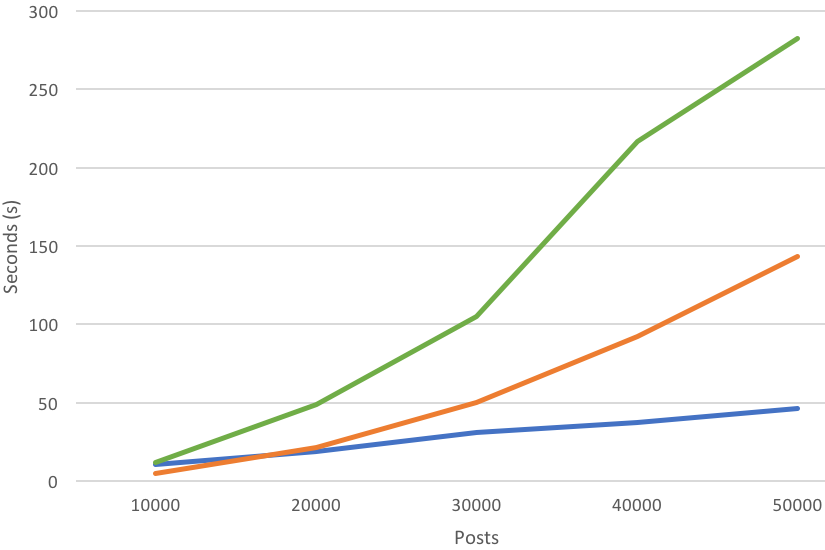}
    \caption{Comparison of run-times for MR-DBSCAN (blue), DBSCAN (red)
      and HDBSCAN (green). Parameter settings are $m_{pts} = 50$ and
      $\varepsilon = 0.01$ for all tests.}
    \label{fig:exp5_res}
\end{figure}

The purpose of this experiment is to compare run-time as a function of
number of forum posts of the three different clustering algorithms
DBSCAN, MR-DBSCAN, and HDBSCAN~\cite{hdbscan_original}. Algorithm
parameters are fixed and equal across the tests in order not to bias
the results. Specifically, the lower point-count threshold for
establishing a core point $m_{pts} = 50$ and the neighborhood radius
$\varepsilon = 0.01$ for all tests. Note that the setting
$\varepsilon = 0.01$ was previously found
(section~\ref{subsection:exp_runtimeanalysis}) to be a suitable choice
for MR-DBSCAN. The data set in this experiment is various subsets of
the collected forum posts; the number of tf-idf features has been
limited to 1000. The results of all tests are reported in
table~\ref{tbl:experiment5_res} and figure~\ref{fig:exp5_res}.

\begin{table}[htbp!]
\setlength\extrarowheight{1pt}
\caption{%
Ten run-time tests of the clustering algorithms MR-DBSCAN, DBSCAN, and HDBSCAN. %
Parameter settings are $m_{pts} = 50$ and $\varepsilon = 0.01$ for all tests. %
Gray cells indicate discontinued tests due to memory exhaustion of the test computer.%
}
\label{tbl:experiment5_res}
\centering
\begin{tabular}{|c|c|c|c|c|}
\hline
\textbf{\#} & \textbf{Posts} & \textbf{MR-DBSCAN [s]} & \textbf{DBSCAN [s]} & \textbf{HDBSCAN [s]}  \\ \hline
\textbf{1} & 10000  & 10.696 & 4.755 & 11.969 \\ \hline
\textbf{2} & 20000 & 18.994 & 21.167 & 48.731 \\ \hline
\textbf{3} & 30000 & 31.115 & 50.237 & 105.007 \\ \hline
\textbf{4} & 40000 & 37.321 & 92.392 & 217.033 \\ \hline
\textbf{5} & 50000 & 46.185 & 143.405 & 282.358 \\ \hline
\textbf{6} & 60000 & 53.235 & \cellcolor{lg} & \cellcolor{lg} \\ \hline
\textbf{7} & 70000 & 66.240 & \cellcolor{lg} & \cellcolor{lg} \\ \hline
\textbf{8} & 80000 & 83.283 & \cellcolor{lg} & \cellcolor{lg} \\ \hline
\textbf{9} & 90000 & 92.393 & \cellcolor{lg} & \cellcolor{lg} \\ \hline
\textbf{10} & 100000 & 113.077 & \cellcolor{lg} & \cellcolor{lg} \\ \hline
\end{tabular}
\end{table}

MR-DBSCAN is slower than DBSCAN in the first test case with 10.000
posts, but from this point on it is executing much faster. When DBSCAN
and HDBSCAN stopped executing due to memory exhaustion of the test
computer, MR-DBSCAN continued; thus, the gray cells in
table~\ref{tbl:experiment5_res} and the x-axis limit in
figure~\ref{fig:exp5_res}. The memory exhaustion when running DBSCAN
and HDBSCAN is mainly due to the growth of the tf-idf matrix which
holds a forum post per row. This is simply not a feasible
implementation when clustering problems become large. Clearly, divide
and conquer by MapReduce help circumvent this problem.

\section{Discussion}
\label{sec:discussion}

We argue that the information hidden in non-clinical texts is valuable
and worth retrieving and activating. In the present study, the
activation is done via a decision support system that helps cancer
patients and caretakers to stay informed about cancer trajectories,
i.e. associated symptoms, diagnoses, treatments, and outcomes, and to make
informed arguments and decisions regarding treatment plans.

Concretely, the presented system analyzes non-clinical forum posts'
contents by using text retrieval, clustering, and classification
methods. The methods are executed in a distributed computing setup,
specifically MapReduce, to achieve computational efficiency via
utilization of multi-cores in modern computers. Indeed, the
computational burdensome clustering was significantly improved in
terms of run-time by MapReduce; thus, the clustering method MR-DBSCAN
is recommendable for large clustering challenges.

Moreover, the presented system provides an interactive graphical user
interface that allows end-users to mine the valuable information and
to get an overview over cancer trajectories. Hopefully, the proposed
system and systems alike will also help build patient/caretaker
communities by leveraging the soft information not hitherto used by
the established health care systems, e.g. information about emotions,
feelings, or personal preferences.

The present study can be extended in several different ways. Adding,
refining, and benchmarking more clustering and classification methods
would yield a more comprehensive comparison that might lead to even
better results, i.e. more accurate clusterings and classifications,
and thus, ultimately, a better end-user service. For the
classification it would especially be of interest to collect and use a
larger training set. With regard to DBSCAN and HDBSCAN clustering, we
experienced memory exhaustion problems on our local development
machines when executing the algorithms on large sets of posts,
i.e. around 50.000 posts. It is of interest to address these memory
consumption challenge by redesigning the algorithms such that upper
bounds on memory consumption can be guaranteed.

Lastly, it would be interesting to generalize the presented system
such that it readily can be applied in other domains besides cancer;
this would require an easy way of loading new data-sets and associated
feature-vectors.

\bmhead{Acknowledgments}

We would like to acknowledge Kim Svendsen (Stibo Accelerator,
Højbjerg, Denmark) and Jacob Høy Berthelsen and Bo Thiesson
(Enversion, Aarhus, Denmark) for inspiring discussions. We would also like
to thank Enversion for providing a web crawler suitable for collecting
cancer forum posts.

\clearpage

\section*{Declarations}

\subsection*{Ethics approval}

Not applicable.

\subsection*{Data availability}

The data that support the findings of this study come from the private Danish company Enversion and is not publicly available. It is however available from the authors upon reasonable request.

\subsection*{Competing interests}

The authors declare that they have no known competing financial interests or personal relationships that could have appeared to influence the work reported in this paper.

\subsection*{Funding}

No funding was received for this study.

\subsection*{Authors' contributions}

J. Agerskov and K. Nielsen conceived the presented idea, implemented the prototype and documented the experiments. C. M. Lillelund assisted with writing the introduction chapter and prepare the manuscript in accordance with Springer guidelines. C. F. Pedersen supervised the project. All authors discussed the results and contributed to the final manuscript.

\bibliography{sn-bibliography}


\begin{thebibliography}{38}
\ifx \bisbn   \undefined \def \bisbn  #1{ISBN #1}\fi
\ifx \binits  \undefined \def \binits#1{#1}\fi
\ifx \bauthor  \undefined \def \bauthor#1{#1}\fi
\ifx \batitle  \undefined \def \batitle#1{#1}\fi
\ifx \bjtitle  \undefined \def \bjtitle#1{#1}\fi
\ifx \bvolume  \undefined \def \bvolume#1{\textbf{#1}}\fi
\ifx \byear  \undefined \def \byear#1{#1}\fi
\ifx \bissue  \undefined \def \bissue#1{#1}\fi
\ifx \bfpage  \undefined \def \bfpage#1{#1}\fi
\ifx \blpage  \undefined \def \blpage #1{#1}\fi
\ifx \burl  \undefined \def \burl#1{\textsf{#1}}\fi
\ifx \doiurl  \undefined \def \doiurl#1{\url{https://doi.org/#1}}\fi
\ifx \betal  \undefined \def \betal{\textit{et al.}}\fi
\ifx \binstitute  \undefined \def \binstitute#1{#1}\fi
\ifx \binstitutionaled  \undefined \def \binstitutionaled#1{#1}\fi
\ifx \bctitle  \undefined \def \bctitle#1{#1}\fi
\ifx \beditor  \undefined \def \beditor#1{#1}\fi
\ifx \bpublisher  \undefined \def \bpublisher#1{#1}\fi
\ifx \bbtitle  \undefined \def \bbtitle#1{#1}\fi
\ifx \bedition  \undefined \def \bedition#1{#1}\fi
\ifx \bseriesno  \undefined \def \bseriesno#1{#1}\fi
\ifx \blocation  \undefined \def \blocation#1{#1}\fi
\ifx \bsertitle  \undefined \def \bsertitle#1{#1}\fi
\ifx \bsnm \undefined \def \bsnm#1{#1}\fi
\ifx \bsuffix \undefined \def \bsuffix#1{#1}\fi
\ifx \bparticle \undefined \def \bparticle#1{#1}\fi
\ifx \barticle \undefined \def \barticle#1{#1}\fi
\bibcommenthead
\ifx \bconfdate \undefined \def \bconfdate #1{#1}\fi
\ifx \botherref \undefined \def \botherref #1{#1}\fi
\ifx \url \undefined \def \url#1{\textsf{#1}}\fi
\ifx \bchapter \undefined \def \bchapter#1{#1}\fi
\ifx \bbook \undefined \def \bbook#1{#1}\fi
\ifx \bcomment \undefined \def \bcomment#1{#1}\fi
\ifx \oauthor \undefined \def \oauthor#1{#1}\fi
\ifx \citeauthoryear \undefined \def \citeauthoryear#1{#1}\fi
\ifx \endbibitem  \undefined \def \endbibitem {}\fi
\ifx \bconflocation  \undefined \def \bconflocation#1{#1}\fi
\ifx \arxivurl  \undefined \def \arxivurl#1{\textsf{#1}}\fi
\csname PreBibitemsHook\endcsname

\bibitem{jensen_2017_1}
\begin{botherref}
\oauthor{\bsnm{Jensen}, \binits{K.}},
\oauthor{\bsnm{Soguero-Ruiz}, \binits{C.}},
\oauthor{\bsnm{Mikalsen}, \binits{K.O.}},
\oauthor{\bsnm{Lindsetmo}, \binits{R.-O.}},
\oauthor{\bsnm{Kouskoumvekaki}, \binits{I.}},
\oauthor{\bsnm{Girolami}, \binits{M.}},
\oauthor{\bsnm{Skrovseth}, \binits{S.O.}},
\oauthor{\bsnm{Augestad}, \binits{K.M.}}:
Analysis of free text in electronic health records for identification of cancer
  patient trajectories.
Scientific Reports
\textbf{7}(46226)
(2017).
\doiurl{10.1038/srep46226}
\end{botherref}
\endbibitem

\bibitem{murray_2005_1}
\begin{botherref}
\oauthor{\bsnm{Murray}, \binits{S.A.}},
\oauthor{\bsnm{Kendall}, \binits{M.}},
\oauthor{\bsnm{Boyd}, \binits{K.}},
\oauthor{\bsnm{Sheikh}, \binits{A.}}:
Illness trajectories and palliative care.
British Medical Journal
\textbf{330}
(2005).
\doiurl{10.1136\%2Fbmj.330.7498.1007}
\end{botherref}
\endbibitem

\bibitem{kraeftens_bekaempelse}
\begin{botherref}
{The Danish Cancer Society}.
\url{www.cancer.dk}.
Accessed: 9-11-2017
\end{botherref}
\endbibitem

\bibitem{Umefjord01042006}
\begin{barticle}
\bauthor{\bsnm{Umefjord}, \binits{G.}},
\bauthor{\bsnm{Hamberg}, \binits{K.}},
\bauthor{\bsnm{Malker}, \binits{H.}},
\bauthor{\bsnm{Petersson}, \binits{G.}}:
\batitle{The use of an internet-based ask the doctor service involving family
  physicians: Evaluation by a web survey}.
\bjtitle{Family Practice}
\bvolume{23}(\bissue{2}),
\bfpage{159}--\blpage{166}
(\byear{2006}).
\doiurl{10.1093/fampra/cmi117}
\end{barticle}
\endbibitem

\bibitem{medical2016consulations}
\begin{barticle}
\bauthor{\bsnm{Umefjord}, \binits{G.}},
\bauthor{\bsnm{Sandstr{\"o}m}, \binits{H.}},
\bauthor{\bsnm{Malker}, \binits{H.}},
\bauthor{\bsnm{Petersson}, \binits{G.}}:
\batitle{Medical text-based consultations on the internet: A 4-year study}.
\bjtitle{International Journal of Medical Informatics}
\bvolume{77}(\bissue{2}),
\bfpage{114}--\blpage{121}
(\byear{2008}).
\doiurl{10.1016/j.ijmedinf.2007.01.009}
\end{barticle}
\endbibitem

\bibitem{Wang2018}
\begin{barticle}
\bauthor{\bsnm{Wang}, \binits{Y.}},
\bauthor{\bsnm{Wang}, \binits{L.}},
\bauthor{\bsnm{Rastegar-Mojarad}, \binits{M.}},
\bauthor{\bsnm{Moon}, \binits{S.}},
\bauthor{\bsnm{Shen}, \binits{F.}},
\bauthor{\bsnm{Afzal}, \binits{N.}},
\bauthor{\bsnm{Liu}, \binits{S.}},
\bauthor{\bsnm{Zeng}, \binits{Y.}},
\bauthor{\bsnm{Mehrabi}, \binits{S.}},
\bauthor{\bsnm{Sohn}, \binits{S.}}, \betal:
\batitle{Clinical information extraction applications: a literature review}.
\bjtitle{Journal of biomedical informatics}
\bvolume{77},
\bfpage{34}--\blpage{49}
(\byear{2018}).
\doiurl{10.1016/j.jbi.2017.11.011}
\end{barticle}
\endbibitem

\bibitem{Luque2019}
\begin{botherref}
\oauthor{\bsnm{Luque}, \binits{C.}},
\oauthor{\bsnm{Luna}, \binits{J.M.}},
\oauthor{\bsnm{Luque}, \binits{M.}},
\oauthor{\bsnm{Ventura}, \binits{S.}}:
An advanced review on text mining in medicine.
WIREs Data Mining and Knowledge Discovery
\textbf{9}(3)
(2019).
\doiurl{10.1002/widm.1302}
\end{botherref}
\endbibitem

\bibitem{Simpson2012}
\begin{barticle}
\bauthor{\bsnm{Simpson}, \binits{M.S.}},
\bauthor{\bsnm{Demner-Fushman}, \binits{D.}}:
\batitle{Biomedical text mining: a survey of recent progress}.
\bjtitle{In: Aggarwal, C., Zhai, C. (eds) Mining Text Data. Springer, Boston,
  MA.}
(\byear{2012}).
\doiurl{10.1007/978-1-4614-3223-4_14}
\end{barticle}
\endbibitem

\bibitem{Meystre2008}
\begin{barticle}
\bauthor{\bsnm{Meystre}, \binits{S.M.}},
\bauthor{\bsnm{Savova}, \binits{G.K.}},
\bauthor{\bsnm{Kipper-Schuler}, \binits{K.C.}},
\bauthor{\bsnm{Hurdle}, \binits{J.F.}}:
\batitle{Extracting information from textual documents in the electronic health
  record: a review of recent research}.
\bjtitle{Yearbook of medical informatics}
\bvolume{17}(\bissue{01}),
\bfpage{128}--\blpage{144}
(\byear{2008}).
\doiurl{10.1055/s-0038-1638592}
\end{barticle}
\endbibitem

\bibitem{Demner-Fushman2009}
\begin{barticle}
\bauthor{\bsnm{Demner-Fushman}, \binits{D.}},
\bauthor{\bsnm{Chapman}, \binits{W.W.}},
\bauthor{\bsnm{McDonald}, \binits{C.J.}}:
\batitle{What can natural language processing do for clinical decision
  support?}
\bjtitle{Journal of biomedical informatics}
\bvolume{42}(\bissue{5}),
\bfpage{760}--\blpage{772}
(\byear{2009}).
\doiurl{10.1016/j.jbi.2009.08.007}
\end{barticle}
\endbibitem

\bibitem{ebadollahi_2010_1}
\begin{bchapter}
\bauthor{\bsnm{Ebadollahi}, \binits{S.}},
\bauthor{\bsnm{Sun}, \binits{J.}},
\bauthor{\bsnm{Gotz}, \binits{D.}},
\bauthor{\bsnm{Hu}, \binits{J.}},
\bauthor{\bsnm{Sow}, \binits{D.}},
\bauthor{\bsnm{Neti}, \binits{C.}}:
\bctitle{Predicting patient's trajectory of physiological data using temporal
  trends in similar patients: {A} system for near-term prognostics}.
In: \bbtitle{Amia Annual Symposium}
(\byear{2010})
\end{bchapter}
\endbibitem

\bibitem{jensen_2014_1}
\begin{botherref}
\oauthor{\bsnm{Jensen}, \binits{A.B.}},
\oauthor{\bsnm{Moseley}, \binits{P.L.}},
\oauthor{\bsnm{Oprea}, \binits{T.I.}},
\oauthor{\bsnm{Elles{\o}e}, \binits{S.G.}},
\oauthor{\bsnm{Eriksson}, \binits{R.}},
\oauthor{\bsnm{Schmock}, \binits{H.}},
\oauthor{\bsnm{Jensen}, \binits{P.B.}},
\oauthor{\bsnm{Jensen}, \binits{L.J.}},
\oauthor{\bsnm{Brunak}, \binits{S.}}:
Temporal disease trajectories condensed from population-wide registry data
  covering 6.2 million patients.
Nature Communications
\textbf{5}(4022)
(2014).
\doiurl{10.1038/ncomms5022}
\end{botherref}
\endbibitem

\bibitem{ji_2016_1}
\begin{botherref}
\oauthor{\bsnm{Ji}, \binits{X.}},
\oauthor{\bsnm{Chun}, \binits{S.A.}},
\oauthor{\bsnm{Geller}, \binits{J.}}:
Predicting comorbid conditions and trajectories using social health records.
IEEE Transactions on NanoBioscience
\textbf{15}(4)
(2016).
\doiurl{10.1109/tnb.2016.2564299}
\end{botherref}
\endbibitem

\bibitem{assale2019}
\begin{barticle}
\bauthor{\bsnm{Assale}, \binits{M.}},
\bauthor{\bsnm{Dui}, \binits{L.G.}},
\bauthor{\bsnm{Cina}, \binits{A.}},
\bauthor{\bsnm{Seveso}, \binits{A.}},
\bauthor{\bsnm{Cabitza}, \binits{F.}}:
\batitle{The revival of the notes field: Leveraging the unstructured content in
  electronic health records}.
\bjtitle{Frontiers in Medicine}
(\byear{2019}).
\doiurl{10.3389\%2Ffmed.2019.00066}
\end{barticle}
\endbibitem

\bibitem{Nehme2021}
\begin{barticle}
\bauthor{\bsnm{Nehme}, \binits{F.}},
\bauthor{\bsnm{Feldman}, \binits{K.}}:
\batitle{Evolving role and future directions of natural language processing in
  gastroenterology}.
\bjtitle{Digestive diseases and sciences}
\bvolume{66}(\bissue{1}),
\bfpage{29}--\blpage{40}
(\byear{2021}).
\doiurl{10.1007/s10620-020-06156-y}
\end{barticle}
\endbibitem

\bibitem{Frunza:2011:MLA}
\begin{barticle}
\bauthor{\bsnm{Frunza}, \binits{O.}},
\bauthor{\bsnm{Inkpen}, \binits{D.}},
\bauthor{\bsnm{Tran}, \binits{T.}}:
\batitle{A machine learning approach for identifying disease-treatment
  relations in short texts}.
\bjtitle{IEEE Transactions on Knowledge and Data Engineering}
\bvolume{23}(\bissue{6}),
\bfpage{801}--\blpage{814}
(\byear{2011}).
\doiurl{10.1109/TKDE.2010.152}
\end{barticle}
\endbibitem

\bibitem{Rosario:2004:CSR:1218955.1219010}
\begin{bchapter}
\bauthor{\bsnm{Rosario}, \binits{B.}},
\bauthor{\bsnm{Hearst}, \binits{M.A.}}:
\bctitle{Classifying semantic relations in bioscience texts}.
In: \bbtitle{ACL, Annual Meeting of the Association for Computational
  Linguistics},
\bconflocation{Stroudsburg, PA, USA}
(\byear{2004}).
\doiurl{10.3115/1218955.1219010}
\end{bchapter}
\endbibitem

\bibitem{yang2011analyzing}
\begin{barticle}
\bauthor{\bsnm{Yang}, \binits{C.C.}},
\bauthor{\bsnm{Ng}, \binits{T.D.}}:
\batitle{Analyzing and visualizing web opinion development and social
  interactions with density-based clustering}.
\bjtitle{IEEE Transactions on Systems, Man, and Cybernetics -- Part A: Systems
  and Humans}
\bvolume{41}(\bissue{6}),
\bfpage{1144}--\blpage{1155}
(\byear{2011}).
\doiurl{10.1109/TSMCA.2011.2113334}
\end{barticle}
\endbibitem

\bibitem{iso_25010}
\begin{botherref}
\oauthor{\bsnm{{ISO/IEC JTC 1/SC 7}}}:
{ISO/IEC} 25010:2011, systems and software engineering -- systems and software
  quality requirements and evaluation (square) -- system and software quality
  models.
Technical Report~1,
{ISO/IEC}
(March 2011).
{ICS 35.080}
\end{botherref}
\endbibitem

\bibitem{csn_cancer}
\begin{botherref}
\oauthor{\bsnm{{American Cancer Society}}}:
{Cancer Survivors Network}.
\url{csn.cancer.org}.
Accessed: 7-11-2017
\end{botherref}
\endbibitem

\bibitem{Introduction_to_information_retrieval}
\begin{bbook}
\bauthor{\bsnm{Manning}, \binits{C.}},
\bauthor{\bsnm{Raghavan}, \binits{P.}},
\bauthor{\bsnm{Schütze}, \binits{H.}}:
\bbtitle{Introduction to Information Retrieval}.
\bpublisher{Cambridge University Press},
\blocation{Cambridge, United Kingdom}
(\byear{2008}).
\doiurl{10.1017/CBO9780511809071}
\end{bbook}
\endbibitem

\bibitem{lovins1968development}
\begin{botherref}
\oauthor{\bsnm{Lovins}, \binits{J.B.}}:
Development of a stemming algorithm.
Technical report,
Electronic Systems Laboratory, MIT,
Cambridge, Massachusetts, USA
(1968)
\end{botherref}
\endbibitem

\bibitem{chris1990another}
\begin{barticle}
\bauthor{\bsnm{Paice}, \binits{C.D.}}:
\batitle{Another stemmer}.
\bjtitle{ACM, SIG on Information Retrieval FORUM}
\bvolume{24}(\bissue{3}),
\bfpage{56}--\blpage{61}
(\byear{1990}).
\doiurl{10.1145/101306.101310}
\end{barticle}
\endbibitem

\bibitem{porter1980algorithm}
\begin{barticle}
\bauthor{\bsnm{Porter}, \binits{M.}}:
\batitle{An algorithm for suffix stripping}.
\bjtitle{Program}
\bvolume{14}(\bissue{3}),
\bfpage{130}--\blpage{137}
(\byear{1980}).
\doiurl{10.1108/eb046814}
\end{barticle}
\endbibitem

\bibitem{simple-proven-approaches-to-text-retrieval}
\begin{botherref}
\oauthor{\bsnm{Robertson}, \binits{S.E.}},
\oauthor{\bsnm{{Sp\"{a}rck Jones}}, \binits{K.}}:
Simple, proven approaches to text retrieval.
Technical Report UCAM-CL-TR-356,
University of Cambridge, Computer Laboratory,
Cambridge, UK
(December 1994)
\end{botherref}
\endbibitem

\bibitem{On_term_selection_for_query_expansion}
\begin{barticle}
\bauthor{\bsnm{Robertson}, \binits{S.}}:
\batitle{On term selection for query expansion}.
\bjtitle{Journal of Documentation}
\bvolume{46}(\bissue{4}),
\bfpage{359}--\blpage{364}
(\byear{1990}).
\doiurl{10.1108/eb026866}
\end{barticle}
\endbibitem

\bibitem{UnderstandingInverseDocument}
\begin{barticle}
\bauthor{\bsnm{Robertson}, \binits{S.}}:
\batitle{Understanding inverse document frequency: On theoretical arguments for
  {IDF}}.
\bjtitle{Journal of Documentation}
\bvolume{60}(\bissue{5}),
\bfpage{503}--\blpage{520}
(\byear{2004}).
\doiurl{10.1108/00220410410560582}
\end{barticle}
\endbibitem

\bibitem{A_statistical_interpretation_of_term_specificity}
\begin{barticle}
\bauthor{\bsnm{Jones}, \binits{K.}}:
\batitle{A statistical interpretation of term specificity and its applications
  in retrieval}.
\bjtitle{Journal of Documentation}
\bvolume{28}(\bissue{1}),
\bfpage{11}--\blpage{21}
(\byear{1972}).
\doiurl{10.1108/eb026526}
\end{barticle}
\endbibitem

\bibitem{Jain2010651}
\begin{barticle}
\bauthor{\bsnm{Jain}, \binits{A.}}:
\batitle{Data clustering: 50 years beyond k-means}.
\bjtitle{Elsevier, Pattern Recognition Letters}
\bvolume{31}(\bissue{8}),
\bfpage{651}--\blpage{666}
(\byear{2010}).
\doiurl{10.1016/j.patrec.2009.09.011}
\end{barticle}
\endbibitem

\bibitem{Campello2013}
\begin{bchapter}
\bauthor{\bsnm{Campello}, \binits{R.}},
\bauthor{\bsnm{Moulavi}, \binits{D.}},
\bauthor{\bsnm{Sander}, \binits{J.}}:
\bctitle{Density-based clustering based on hierarchical density estimates}.
In: \bbtitle{PAKDD, Advances in Knowledge Discovery and Data Mining},
pp. \bfpage{160}--\blpage{172}.
\bpublisher{Springer},
\blocation{Gold Coast, Australia}
(\byear{2013}).
\doiurl{10.1007/978-3-642-37456-2_14}
\end{bchapter}
\endbibitem

\bibitem{data_mining_concepts}
\begin{bbook}
\bauthor{\bsnm{Han}, \binits{J.}},
\bauthor{\bsnm{Kamber}, \binits{M.}},
\bauthor{\bsnm{Pei}, \binits{J.}}:
\bbtitle{Data Mining: Concepts and Techniques},
\bedition{3}rd edn.
\bpublisher{Morgan Kaufmann Publishers},
\blocation{Burlington, Massachusetts, USA}
(\byear{2011}).
\doiurl{10.1016/C2009-0-61819-5}
\end{bbook}
\endbibitem

\bibitem{mrdbscan2011}
\begin{bchapter}
\bauthor{\bsnm{He}, \binits{Y.}},
\bauthor{\bsnm{Tan}, \binits{H.}},
\bauthor{\bsnm{Luo}, \binits{W.}},
\bauthor{\bsnm{Mao}, \binits{H.}},
\bauthor{\bsnm{Ma}, \binits{D.}},
\bauthor{\bsnm{Feng}, \binits{S.}},
\bauthor{\bsnm{Fan}, \binits{J.}}:
\bctitle{{MR-DBSCAN}: An efficient parallel density-based clustering algorithm
  using {MapReduce}}.
In: \bbtitle{IEEE, International Conference on Parallel and Distributed
  Systems},
\bconflocation{Tainan, Taiwan},
pp. \bfpage{473}--\blpage{480}
(\byear{2011}).
\doiurl{10.1109/ICPADS.2011.83}
\end{bchapter}
\endbibitem

\bibitem{Ester96adensity_based}
\begin{bchapter}
\bauthor{\bsnm{Ester}, \binits{M.}},
\bauthor{\bsnm{Kriegel}, \binits{H.-P.}},
\bauthor{\bsnm{Sander}, \binits{J.}},
\bauthor{\bsnm{Xu}, \binits{X.}}:
\bctitle{A density-based algorithm for discovering clusters in large spatial
  databases with noise}.
In: \bbtitle{ACM, Conference on Knowledge Discovery and Data Mining (KDD)},
pp. \bfpage{226}--\blpage{231}.
\bpublisher{{AAAI Press}},
\blocation{Portland, Oregon, USA}
(\byear{1996})
\end{bchapter}
\endbibitem

\bibitem{xu1999fast}
\begin{barticle}
\bauthor{\bsnm{Xu}, \binits{X.}},
\bauthor{\bsnm{J{\"a}ger}, \binits{J.}},
\bauthor{\bsnm{Kriegel}, \binits{H.-P.}}:
\batitle{A fast parallel clustering algorithm for large spatial databases}.
\bjtitle{Data Mining and Knowledge Discovery}
\bvolume{3}(\bissue{3}),
\bfpage{263}--\blpage{290}
(\byear{1999}).
\doiurl{10.1023/A:1009884809343}
\end{barticle}
\endbibitem

\bibitem{mrdbscan:He2014}
\begin{barticle}
\bauthor{\bsnm{He}, \binits{Y.}},
\bauthor{\bsnm{Tan}, \binits{H.}},
\bauthor{\bsnm{Luo}, \binits{W.}},
\bauthor{\bsnm{Feng}, \binits{S.}},
\bauthor{\bsnm{Fan}, \binits{J.}}:
\batitle{{MR-DBSCAN}: A scalable {MapReduce}-based {DBSCAN} algorithm for
  heavily skewed data}.
\bjtitle{Springer, Frontiers of Computer Science}
\bvolume{8}(\bissue{1}),
\bfpage{83}--\blpage{99}
(\byear{2014}).
\doiurl{10.1007/s11704-013-3158-3}
\end{barticle}
\endbibitem

\bibitem{training_set_bio}
\begin{botherref}
Data for research on relations between DISEASE/TREATMENT entities.
\url{http://biotext.berkeley.edu/dis_treat_data.html}.
Accessed: 16-11-2017
\end{botherref}
\endbibitem

\bibitem{Hubert1985}
\begin{barticle}
\bauthor{\bsnm{Hubert}, \binits{L.}},
\bauthor{\bsnm{Arabie}, \binits{P.}}:
\batitle{Comparing partitions}.
\bjtitle{Springer, Journal of Classification}
\bvolume{2}(\bissue{1}),
\bfpage{193}--\blpage{218}
(\byear{1985}).
\doiurl{10.1007/BF01908075}
\end{barticle}
\endbibitem

\bibitem{hdbscan_original}
\begin{botherref}
\oauthor{\bsnm{Campello}, \binits{R.}},
\oauthor{\bsnm{Moulavi}, \binits{D.}},
\oauthor{\bsnm{Zimek}, \binits{A.}},
\oauthor{\bsnm{Sander}, \binits{J.}}:
Hierarchical density estimates for data clustering, visualization, and outlier
  detection.
ACM Transactions on Knowledge Discovery from Data
\textbf{10}(1)
(2015).
\doiurl{10.1145/2733381}
\end{botherref}
\endbibitem

\end{thebibliography}

\end{document}